\documentclass[aps,prd,showpacs,amssymb,superscriptaddress, notitlepage,floats,floatfix,nofootinbib, twocolumn]{revtex4-1}
\usepackage{graphicx}
\usepackage{amssymb}
\usepackage{amsmath}
\usepackage{amsfonts}

\usepackage{bm}
\usepackage{mathrsfs}
\usepackage[colorlinks=true]{hyperref}
\usepackage[all]{hypcap} 
\usepackage[utf8]{inputenc} 
\usepackage{mathptmx}      % use Times fonts if available on your TeX system
\usepackage{color}
\usepackage{multirow}
\usepackage{palatino}
\usepackage{float}
\usepackage{journals}
\usepackage{soul}
\usepackage{calrsfs}
\DeclareMathAlphabet{\pazocal}{OMS}{zplm}{m}{n}

\begin{document}
%%%%%%%%%%%%%%%%%%%%%%%%%%%%%%%%%%
% % %
\title{
Scalar Fields and Parametrized Spherically Symmetric Black Holes: 
\\Can one hear the shape of space-time?
}

\author{Sebastian H. V\"olkel }
\email{sebastian.voelkel@uni-tuebingen.de}
\author{Kostas D. Kokkotas}%\thanks{kostas.kokkotas@uni-tuebingen.de}}
\affiliation{Theoretical Astrophysics, IAAT, University of T\"ubingen, Germany}
\date{\today}
% % % 
%%%%%%%%%%%%%%%%%%%%%%%%%%%%%%%%%%
\begin{abstract}
In this work we study whether parametrized spherically symmetric black hole solutions in metric theories of gravity can appear to be isospectral when studying perturbations. From a theory agnostic point of view, the test scalar field wave equation is the ideal starting point to approach the quasi-normal mode spectrum in alternative black hole solutions. We use a parametrization for the metric proposed by Rezzolla and Zhidenko, as well as the the higher order WKB method in the determination of the quasi-normal mode spectra. We look for possible degeneracies in a tractable subset of the parameter space with respect to the Schwarzschild quasi-normal modes. Considering the frequencies and damping times of the expected observationally most relevant quasi-normal modes, we find such degeneracies. We explicitly demonstrate that the leading Schwarzschild quasi-normal modes can be approximated by alternative black hole solutions when their mass is treated as free parameter. In practice, we conclude that the mass has to be known with extremely high precision in order to restrict the leading terms in the metric expansion to currently known limits coming from the PPN expansion. Possible limitations of using the quasi-normal mode ringdown to investigate black hole space-times are being discussed.
\end{abstract}
\maketitle
\section{Introduction}
In Einstein's theory of general relativity \cite{einstein}, the field equations determine the metric tensor, which is related to the present energy and matter. With the metric tensor at hand, different types of perturbations on this background can be studied and eventually lead to quasi-normal mode spectra, see \cite{1999LRR.....2....2K,1999CQGra..16R.159N,2009CQGra..26p3001B} for some reviews. The quasi-normal mode spectrum of a certain type of perturbation is an important and characteristic property connecting linear oscillation phenomena with compact objects, e.g. gravitational perturbations of black holes \cite{1957PhRv..108.1063R,1970PhRvL..24..737Z,1973ApJ...185..635T} and neutron stars \cite{1967ApJ...149..591T}. The simplest perturbation is a scalar test field, described by the massless Klein-Gordon equation. It is the simplest example, because once the background metric is known, there is no direct connection to the rather complicated field equations of general relativity. This is different for gravitational perturbations, which are obtained by expanding the field equations in terms of a small perturbation of the metric tensor. In this case, even at linearized order, the field equations determine how perturbations evolve. The importance of this key difference becomes crucial if one starts to be interested in alternative metric theories of gravity. Although general relativity has been passing all tests for more than 100 years \cite{Will:2014kxa}, the strong field regime around black holes only became accessible recently, via gravitational waves e.g. \cite{LIGO1,LIGO2} and black hole shadows \cite{Akiyama:2019cqa}.
\par
It is also possible that different field equations admit a subclass of similar solutions, see \cite{Sotiriou_2015} for a discussion on scalar tensor theories and \cite{Suvorov:2019qow} for the gravitational perturbations of the Kerr solution in $f(R)$ theories. However, since the field equations differ, the evolution of gravitational perturbations will in general be different. This implies that even if a metric tensor is provided or assumed, it is necessary to know the underlying theory if one is interested in gravitational waves. This is in strong contrast to the test scalar field evolving independent of the field equations. Of course, one could also imagine the rather exotic case that two alternative theories yield slightly different solutions for the background, but evolve perturbations in such a way that the observations are too similar to be distinguishable. Thus, if one is agnostic about the underlying theory of gravity, but interested in describing gravitational waves, one faces a fundamental problem.
\par
In general relativity, scalar test fields behave qualitatively to a large extend similar to gravitational perturbations. In the Schwarzschild space-time \cite{1916SPAW.......189S}, scalar and gravitational perturbations are both described by an effective one-dimensional wave equation that includes a potential term \cite{1957PhRv..108.1063R,1970PhRvL..24..737Z}. The explicit form of the potential differs quantitatively, but is qualitatively very similar. Moreover, by going to the Eikonal limit (large $l$ limit), both potentials approach each other. Although such similarities can clearly not be expected in every alternative theory, it is still reasonable that they are encoded in some of them. Therefore, as long as one is aware of this circumstance and restricted to such theories, this connection is a promising starting point to study the qualitative gravitational wave behavior in alternative black hole solutions. 
\par
Instead of picking a few explicit examples from specific theories, we here are interested in parametrized black hole metrics. Such a parametrization might or might not be an exact solution in a known or unknown theory. In any case, it can be an effective approximation and therefore serves as useful tool in explicit calculations. Different parameterizations for non-rotating, as well as rotating, black hole solutions have been proposed in the literature, e.g. \cite{Johannsen:2011dh,PhysRevD.83.104027,PhysRevD.90.084009,Konoplya:2016jvv,Papadopoulos:2018nvd}. The idea is to introduce extra terms that can capture possible deviations from the Kerr solution \cite{PhysRevLett.11.237}. Following such approaches can be promising even beyond oscillation phenomena, because they can also be applied to the geodesic motion of test particles, the determination of black hole shadows \cite{Mizuno:2018lxz}, and x-ray reflection spectroscopy \cite{Bambi_2017,Nampalliwar:2019iti}. However, it is clear that not every parametrization will be a reasonable physical solution and pathologies might arise.
\par
In this work we use the proposal by Rezzolla and Zhidenko \cite{PhysRevD.90.084009}, which can be used for non-rotating black hole solutions. The advantage of this parametrization is that a small number of parameters can be used to describe deviations from the Schwarzschild solution in an explicit way that incorporates proper conditions at spatial infinity and at the horizon. Moreover, it was demonstrated \cite{PhysRevD.90.084009,Kokkotas:2017zwt} that this proposal allows for a precise and fast converging approximation for exact solutions in alternative theories.
\par
With this background information at hand, and some inspiration from Marc Kac's question whether one can hear the shape of a drum \cite{10.2307/2313748}, the motivation of our work is the following. From a theory and metric agnostic point of view, one can ask the following question:
\par
\textit{How uniquely is the quasi-normal mode spectrum related to the theory and metric under consideration?}
\par
Since the answer in the case of gravitational perturbations involves the underlying field equations, it can not be answered explicitly. Connecting quasi-normal mode spectra with compact objects has lead to the field of black hole spectroscopy \cite{Dreyer_2004}. Now, if one makes the assumption that the underlying field equations are very similar, and only introduces changes in the metric, a study on the time evolution of axial perturbations has been done in \cite{Pei:2015sqa}. A perturbative framework, which relates small, parametrized deviations from different perturbation potentials to their quasi-normal mode spectrum, has been proposed in \cite{PhysRevD.99.104077} and extended to coupled wave equations \cite{ringdown2} arising in alternative theories of gravity. Related approaches using the Eikonal limit can be found in \cite{PhysRevD.96.064054,Glampedakis:2019dqh}.
\par
As another approach, at least for scalar test fields, one could start with an efficient parametrization for different space-times and study the related eigenvalue problem. Even on the level of scalar fields, there is a rich complexity that can arise in the possible black hole space-times. Therefore, it is a priori not trivial to answer whether the same quasi-normal mode spectrum can originate from different space-times.
\par
Since the complete quasi-normal mode spectrum is infinite in general relativity, one can also rephrase the question in a more observation driven context: 
\par
\textit{Assuming that observations would provide the first few quasi-normal modes within a finite error, how likely is it that the reconstructed parameters determine these eigenvalues uniquely?}
\par
It is clear that the underlying theory of gravity is needed to describe actual gravitational wave observations. However, test scalar fields will at least in some theories behave qualitatively similar and do not require more than the background space-time. Therefore, we consider a scalar test field evolving in the background of the parametrized Rezzolla-Zhidenko metric as a theory agnostic way to study the qualitative behavior of fields and their quasi-normal mode spectra.
\par
We are particularly interested in the quasi-normal mode properties, because at least for moderate deviations from Schwarzschild, they are strongly influenced by the metric around the lightring ($\sim 3\,M$). This is much closer to the horizon than accretion discs and therefore potentially more sensitive to deviations.
\par
This paper is organized as follows. In Sec. \ref{background} we briefly outline the Rezzolla-Zhidenko parametrization along with the scalar wave equation. Our framework for the study of the quasi-normal mode spectrum of different black hole space-times is explained in Sec. \ref{setup}. Our results are then presented in Sec. \ref{results} and discussed in Sec. \ref{discussion}. Finally, we conclude in Sec. \ref{conclusions}. Throughout the paper we use $G=c=1$.
\section{Black Hole Parametrization and Scalar Perturbations}\label{background}
The starting point in this work is the Rezzolla-Zhidenko (RZ) parametrization for spherically symmetric black hole metrics \cite{PhysRevD.90.084009}. Instead of reproducing all the details, we only give a short summary and refer to the original work for detailed discussion and applications.
\subsection{Parametrization of the RZ Metric}\label{parametrization of the metric}
The line element for spherically symmetric black holes described by the RZ metric takes the form
\begin{align}
\text{d}s^2 = -N^2(r) \text{d}t^2 + \frac{B^2(r)}{N^2(r)} \text{d}r^2 + r^2 \text{d} \Omega^2,
\end{align}
with $\text{d}\Omega^2 = \text{d}\theta^2 + \sin^2 \theta \text{d}\phi^2$. All details of the space-time are encoded in the two functions $N(r)$ and $B(r)$. It turns out to be useful to introduce the dimensionless variable
\begin{align}
x \equiv 1-\frac{r_0}{r},
\end{align}
which maps the location of the event horizon $r_0$ to $x=0$ and spatial infinity to $x=1$. Another function $A(x)$ is introduced via
\begin{align}
N^2 = x A(x),
\end{align}
with $0 < A(x)$ for $0 \leq x \leq 1$. The parametrization will in practice be described by the expansion coefficients that determine $A(x)$ and $B(x)$, which are defined by
\begin{align}
A(x) &= 1 - \varepsilon(1-x) + (a_0 -\epsilon)(1-x)^2 + \tilde{A}(x)(1-x)^3,
\\
B(x) &= 1+ b_0(1-x) + \tilde{B}(x)(1-x)^2.
\end{align} 
The functions $\tilde{A}(x)$ and $\tilde{B}(x)$ should describe the metric near the horizon and at spatial infinity. A connection to the PPN expansion allows to constrain their behavior around $x\approx 1$ and relate them to the PPN parameters $\beta, \gamma$. In the same work it was shown that
\begin{align}
\epsilon &= - \left(1-\frac{2M}{r_0} \right),
\\
a_0&=\frac{(\beta-\gamma)(1+\varepsilon)^2}{2},
\\
b_0&= \frac{(\gamma-1)(1+\varepsilon)}{2}.
\end{align}
By using the experimentally constrained values for $\beta$ and $\gamma$, it follows that $a_0 \sim b_0 \sim 10^{-4}$. Finally, the functions $\tilde{A}(x)$ and $\tilde{B}(x)$ are expressed with their Pad\'e approximants as continued fractions
\begin{align}
\tilde{A}(x) &= \frac{a_1}{1+\frac{a_2x}{1+\frac{a_3x}{1+\dots}}},
\\
\tilde{B}(x) &= \frac{b_1}{1+\frac{b_2x}{1+\frac{b_3x}{1+\dots}}}.
\end{align}
\subsection{Scalar Wave Equation}
The simplest, but also gravitational theory independent wave equation is the Klein-Gordon equation. For spherically symmetric space-times it is well known that it reduces effectively to solve
\begin{align}\label{wave-equation}
\frac{\text{d}^2}{\text{d}{r^{*}}^2}\Psi + \left[\omega_n^2 - V_l(r) \right] \Psi  = 0,
\end{align}
where $r^{*}$ is the tortoise coordinate related to the parametrized metric via
\begin{align}
\frac{\text{d} r^{*}}{\text{d}r} = \frac{B(r)}{N^2(r)},
%\sqrt{\frac{g_{11}}{g_{00}}},
\end{align}
and the effective potential $V_l(r)$ is given by
\begin{align}
V_l(r) = \frac{l(l+1)}{r^2}N^2(r) + \frac{1}{r} \frac{\text{d}}{\text{d}r^{*}} \frac{N^2(r)}{B(r)}.
\end{align}
Note that the simple time independent Schr\"odinger form in eq. \eqref{wave-equation} can in practice be tricky, because the explicit determination of the tortoise coordinate $r^*(r)$, as well as its inversion $r(r^*)$, might in general not be possible with known functions. This makes the explicit evaluation of $V_l(r(r^*))$ a simple numerical exercise. However, this form of the wave equation has the major benefit that many well studied methods to solve it exist. 
\par
The shape of $V_l(r)$ of the Schwarzschild black hole is the one of a two turning point potential barrier, with a maximum close to the lightring at $3\,M$. In a series of seminal works, it was shown how the WKB method can be used to derive an extremely useful explicit solution to the eigenvalue problem of outgoing waves at infinity and ingoing waves at the horizon of different types of black holes \cite{1985ApJ...291L..33S,1987PhRvD..35.3621I,PhysRevD.35.3632,PhysRevD.37.3378,PhysRevD.41.374,Kokkotas_1991,PhysRevD.68.024018}.
\par
In the non-rotating case, the quasi-normal mode spectrum $\omega_n$ is determined by
\begin{align}\label{wkb-expansion}
\frac{i Q_0}{\sqrt{2 Q^{\prime \prime}_0}} - \Lambda_2 - \Lambda_3 - \Lambda_4 - \Lambda_5 - \Lambda_6 = n + \frac{1}{2},
\end{align}
with $Q(r^{*}) \equiv \omega_n^2 - V_l(r^{*})$, which is evaluated at the maximum of potential \cite{PhysRevD.68.024018}. The primes denote the derivative with respect to the tortoise coordinate, also evaluated at the maximum. Without the correction terms $\Lambda_i$, the parabolic result of Mashhoon \cite{1983mgm..conf..599M}\footnote{Using the bound states of the inverted parabolic potential.}, and Schutz and Will is recovered \cite{1985ApJ...291L..33S}\footnote{Using the parabolic barrier explicitly.}. The index of the correction terms show the order in the WKB expansion and are explicitly determined by the overtone number $n$, as well as derivatives of the potential with respect to the tortoise coordinate, evaluated at the maximum of potential. Since their explicit form is rather long, we refer to \cite{PhysRevD.68.024018}.
\par
The main advantage of eq. \eqref{wkb-expansion} is that all terms can be evaluated, once the potential and the tortoise transformation are being provided. The method is expected to give accurate results for $n \leq l$. The overall performance improves with increasing $l$ and decreases with increasing $n$. As long as the general criteria for the validity of the WKB method are fulfilled \cite{1978amms.book.....B}, and the potential barrier has two turning points, the method provides a quick and robust determination of the leading quasi-normal modes. Luckily, these are the ones which are dominant in the ringdown phase and therefore of key importance from an observational point of view.
\section{Search for Isospectrality}\label{setup}
The purpose of our work is to study how the quasi-normal mode spectrum and the parametrized metric are connected to each other. More specifically, we want to investigate whether different metrics can yield the same quasi-normal mode spectrum. This would imply that the effective potentials are either exactly the same or differ, but are still isospectral. At the same time, we also want to relate the question to the situation that only some part of the quasi-normal mode spectrum is provided, but with finite errors. In this case we then want to know if a degeneracy in the parameter space of alternative black hole solutions with respect to the Schwarzschild solution can arise. 
\par
In order to  address both questions, we find ourselves at the crossroads of following a more mathematical inspired path or a more applied one. By mathematical we mean e.g. to apply the Darboux transform, as it was recently done in \cite{PhysRevD.96.024036}, to confirm the already known isospectality of the Regge-Wheeler and Zerilli potentials of gravitational perturbations in general relativity. The more applied path we chose is to start with a finite set of parameters that describe the metric and search in the parameter space for degeneracy by solving the eigenvalue problem explicitly for reasonable combination of parameters.
\par
In the following we outline our approach to study the parameter space of different metrics and how we relate it with a given spectrum.
\subsection{Provided Spectrum}
In order to study isospectrality, one needs to define a potential or a spectrum as reference first. The canonic choice for a spherically symmetric black hole is the Schwarzschild solution \cite{1916SPAW.......189S}. We will use its quasi-normal mode spectrum as reference and define the relative error for the real (r) and imaginary (i) parts of $\omega$ as
\begin{align}\label{relative_error}
\delta \omega_\text{r,i} \equiv \frac{\omega^\text{Sch}_\text{r,i} - \omega^\text{param}_\text{r,i}}{\omega^\text{Sch}_\text{r,i}},
\end{align}
where the explicit choice of $n$ and $l$, as well as the parameters are suppressed. In order to combine the knowledge of the real and imaginary part of a given mode, we also define a combined relative error as
\begin{align}\label{combined_relative_error}
\delta \omega_\text{c} \equiv \sqrt{\delta \omega_\text{r}^2 + \delta \omega_\text{i}^2},
\end{align}
where $\delta \omega_\text{r,i}$ are the standard relative errors defined in eq. \eqref{relative_error}.
\par
A parametrized potential is isospectral to the Schwarzschild potential, if $\delta \omega_\text{r,i}=0$, for all $l$ and $n$. Since there are infinitely many eigenvalues, the full problem can in practice not be studied numerically. However, coming back to the connection to observations, one can only measure a finite set of eigenvalues. A reasonable choice is to start with the knowledge of the fundamental quasi-normal modes for moderate $l$ and include a finite set of eigenvalues by increasing $l$ and $n$. One can then compute the relative errors explicitly in the parameter space. Since the quasi-normal mode spectrum is complex valued, it is also interesting to see how its real and imaginary parts separately depend on the parameter space, as well as their combination.
\subsection{Studied Parameter Space}\label{parameter_range}
In \cite{PhysRevD.90.084009} it was demonstrated that the parametrized metric has very good convergence properties with respect to the number of included parameters. Thus, even including a relatively small set of parameters can yield accurate results, which implies that many alternative black hole solutions are being included. This is an important observation for our work. A large set of free parameters makes it likely that a small set of observables can more easily be explained than if the set of free parameters is small and therefore less favored.
\par
The whole parameter space of the most general metric is infinitely large. However, since we are only interested in those space-times that can be well approximated with a few terms, we consider a finite parameter space. As such, we choose the ADM mass $M$ and only the leading terms for $[a_i,b_i]$
\begin{align}\label{parameters}
\pazocal{P} \equiv [M, a_0, a_1, b_0, b_1],
\end{align}
while all other parameters $[\varepsilon, a_i, b_i]$, with $2 \leq i$ are set to zero. The deviation of the horizon location, described by $\varepsilon$, is set to zero as well. Allowing for more non-zero parameters is in principle interesting as well, but the number of combinations becomes less tractable. We set $\varepsilon$ to zero in this work, because it appears in the lowest order of the expansion and the next higher order terms $[a_0, b_0]$ are already constrained to be small.
\par
Note that not all actual values for the parameters can be chosen arbitrarily. Naturally the ADM mass $M$ is positive. Furthermore, in order to keep the signature of the metric, the expansion parameters describing $\tilde{A}(x)$ have to be chosen such that $0 < A(x)$, for $0\leq x \leq 1$. Also, since we are using the WKB method, we have to make sure that all effective potentials in the parameter space being considered are actually two turning point potential barriers.
\subsection{Higher Order WKB and Numerical Implementation}
For every choice of the parameters $\pazocal{P}$ one has to solve the eigenvalue problem. In practice we are interested in the leading quasi-normal modes with moderate $l$ and $n < l$. First, these modes are the most relevant ones from an hypothetical observational point of view. Second, for these modes it is justified to use the WKB method for numerical results. Applying the WKB method is in principle trivial, but there are some practical aspects we want to note. Obviously, it gets computationally expensive, once the parameter space is investigated with high resolution. For a given computational time, there is a clear trade-off between the quasi-normal mode precision, by including more higher order WKB terms $\Lambda_i$, and the resolution of the parameter space $\pazocal{P}$. The reason being that going one order higher in the WKB method requires two more higher order derivatives of the potential, evaluated in the tortoise coordinate. In the case of a general parametrized metric, this becomes more and more involved. At the same time, keeping a smaller number of WKB terms yields slightly less precise numerical results, but allows to study the parameter space in more detail.
\par
For the explicit numerical calculations we wrote a code. The derivatives of the potential at the maximum are computed with finite differences. Here we avoid the explicit computation of the tortoise coordinate and instead use an iterative scheme
\begin{align}
\frac{\text{d}}{\text{d}r^*}= \frac{N^2(r)}{B(r)} \frac{\text{d}}{\text{d}r},
\end{align}
to obtain the derivative of the potential. The advantage is that we reduce numerical errors arising from the coordinate transformation, but pay the price that higher order derivatives become numerically more expensive. In principle computer algebra software like Maple or Mathematica can compute the derivatives analytically, but the number of involved terms grows rapidly, limiting the overall performance when studying a large parameter space with high WKB order. We choose the third order WKB method for the reasons being mentioned.
\section{Results}\label{results}
In the following we discuss our results for the studied parameter space $\pazocal{P}$ eq. \eqref{parameters}. We consider the fundamental modes ($n=0$) , as well as the first overtone $(n=1)$, for $l=2$ and $l=3$, to make the qualitative connections to gravitational perturbations. This section is organized in different parts. In Sec. \ref{results-1} we compare the properties of the quasi-normal mode spectrum by varying $[a_0, b_0]$ and $[a_1, b_1]$ with respect to the Schwarzschild case. In Sec. \ref{results-2} we show what precision in the black hole mass $M$ is required, in order to determine $ [a_0,b_0]$ and how this depends on the number of provided modes. We do the same analysis for $[a_1, b_1]$ in Sec. \ref{results-3}. We discuss how these results change if one considers the first overtone $(n=1)$ in Sec. \ref{results-4}. Finally, we show some of the reconstructed potentials in Sec. \ref{effective potentials}. Note that parameters not being explicitly varied are set to zero.
\subsection{Parameters $[a_0,b_0]$ and $ [a_1,b_1]$}\label{results-1}
\begin{figure*}[ht]
	\centering
	\includegraphics[width=0.35\linewidth]{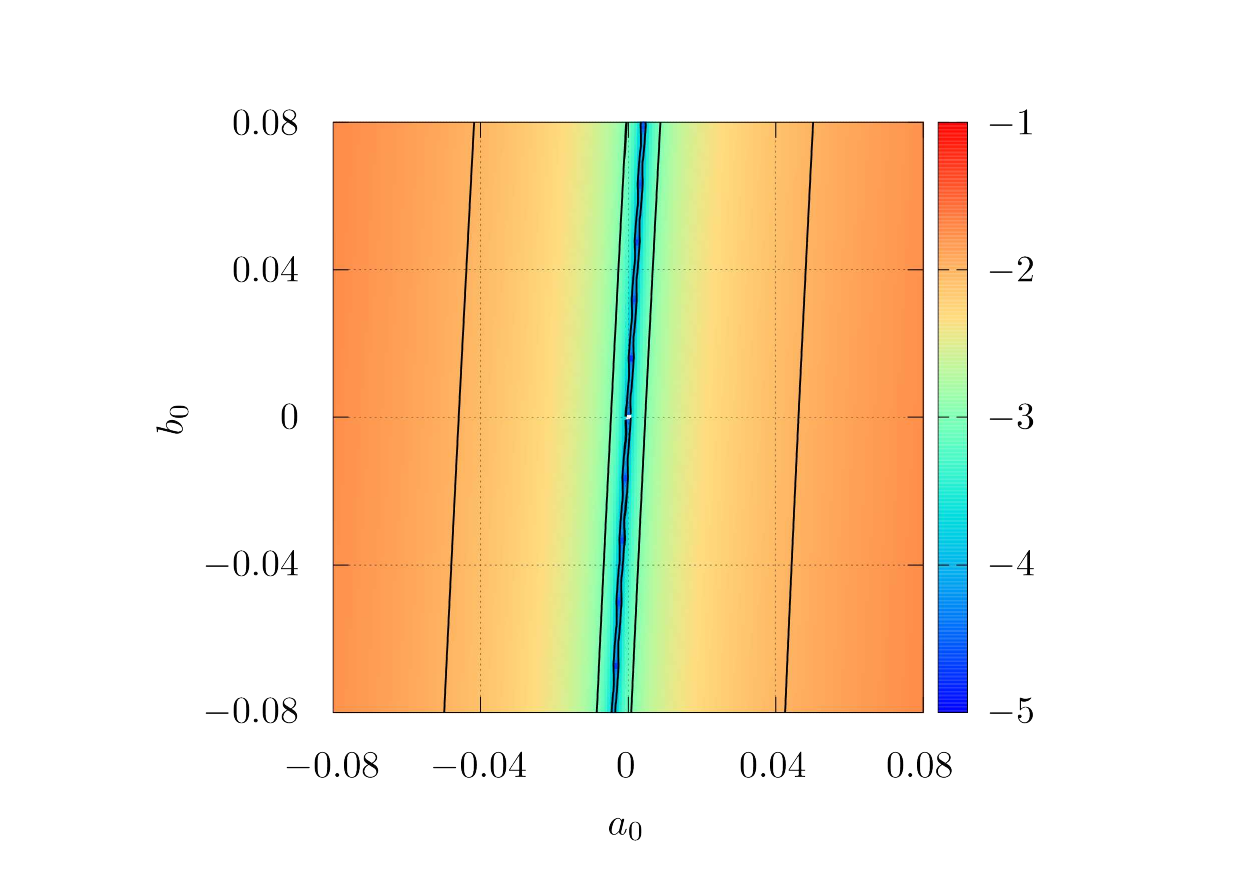}
	\hspace{-0.08\linewidth}
	\includegraphics[width=0.35\linewidth]{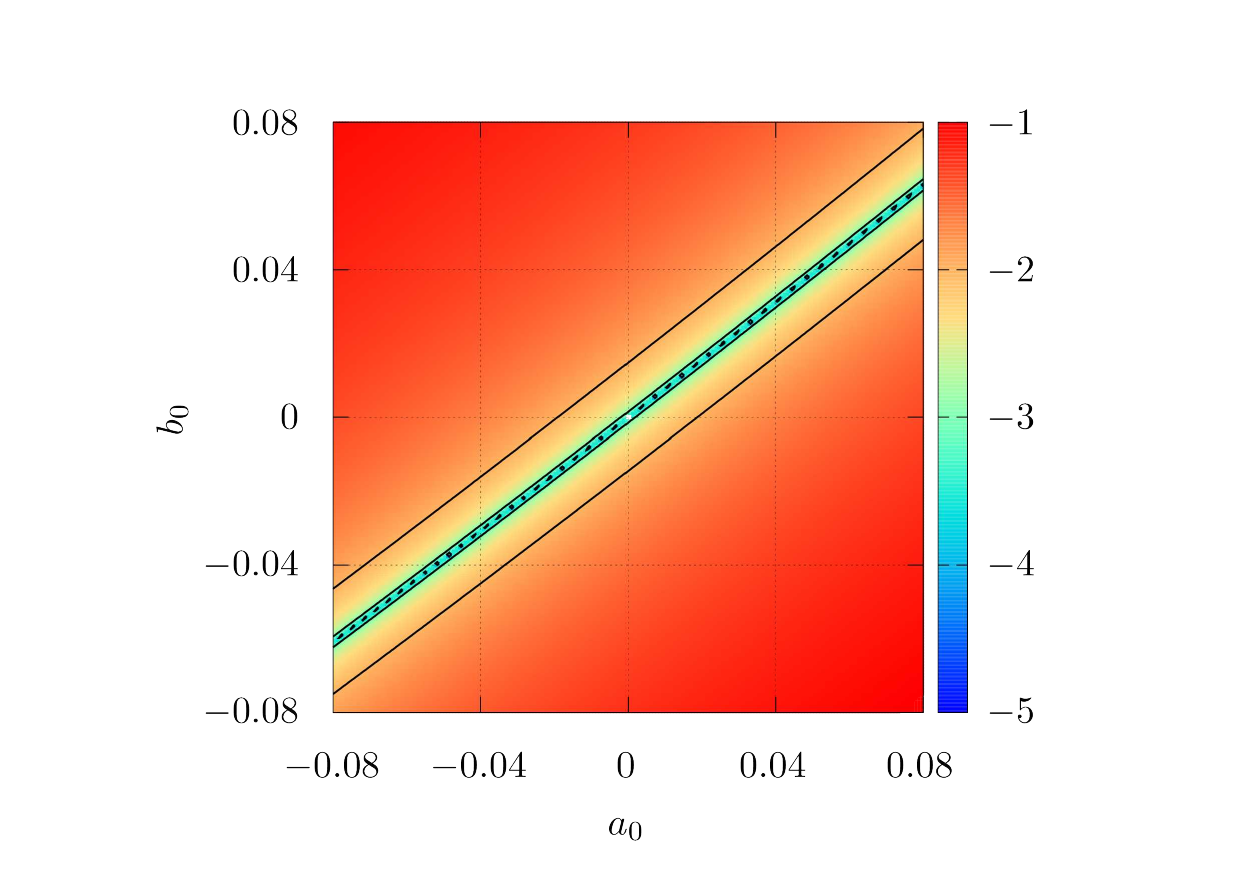}
	\hspace{-0.08\linewidth}
	\includegraphics[width=0.35\linewidth]{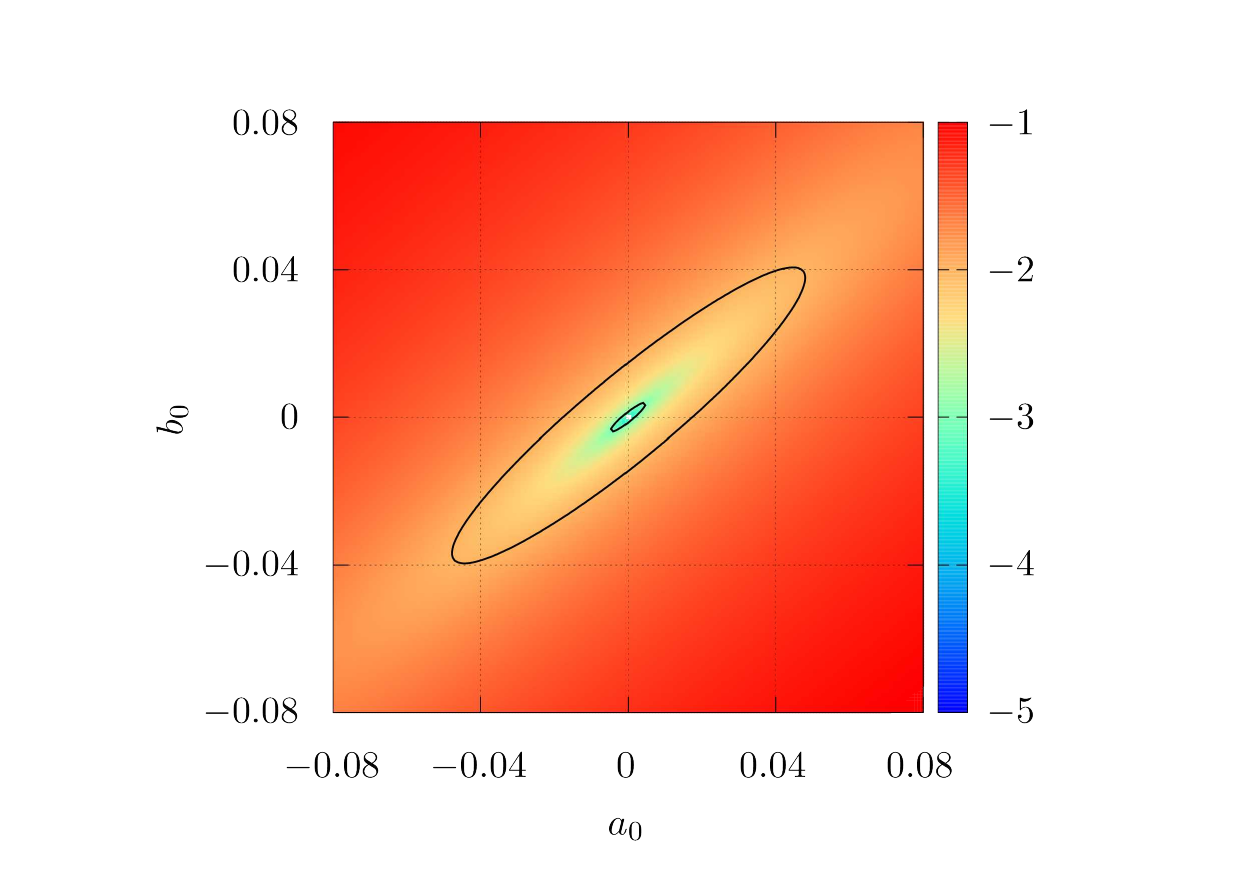}
	\\
	\centering
	\includegraphics[width=0.35\linewidth]{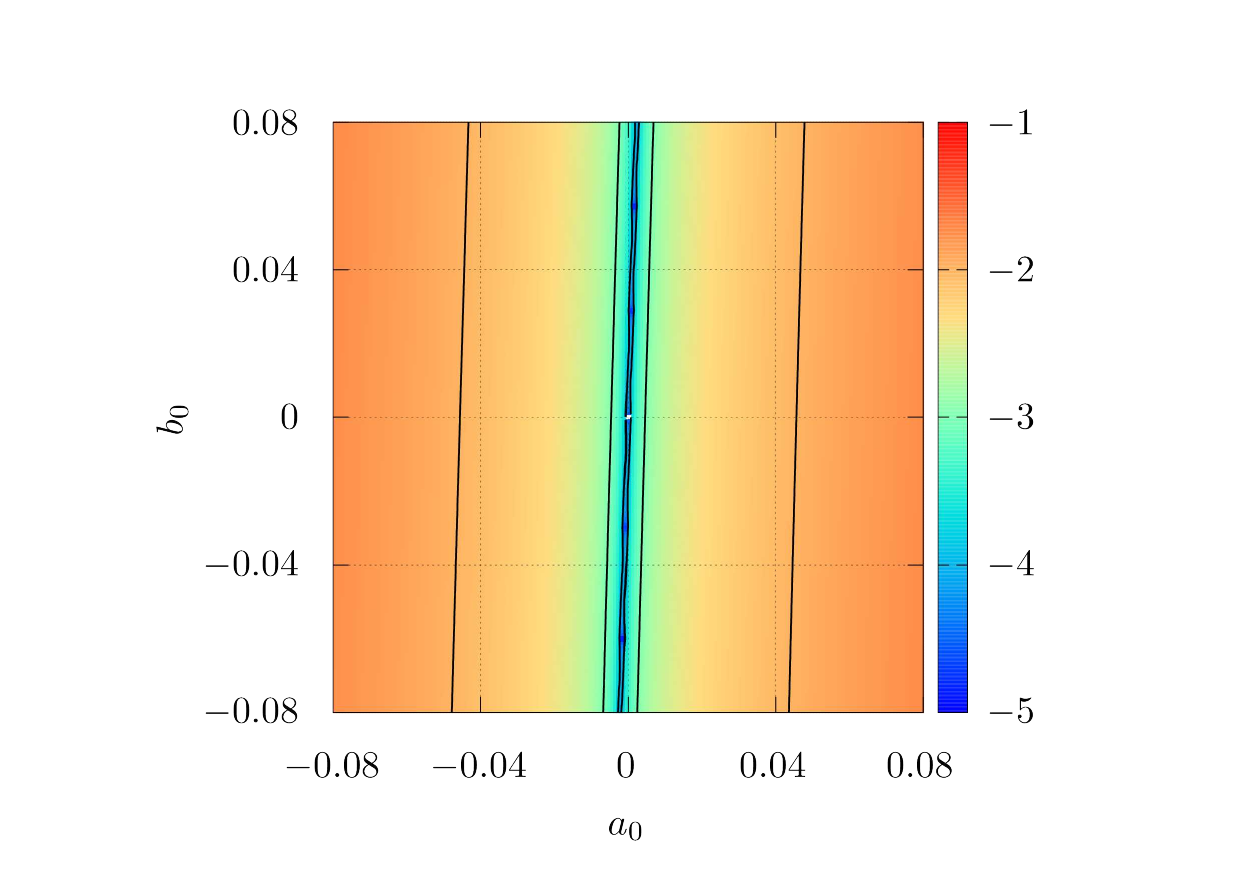}
	\hspace{-0.08\linewidth}
	\includegraphics[width=0.35\linewidth]{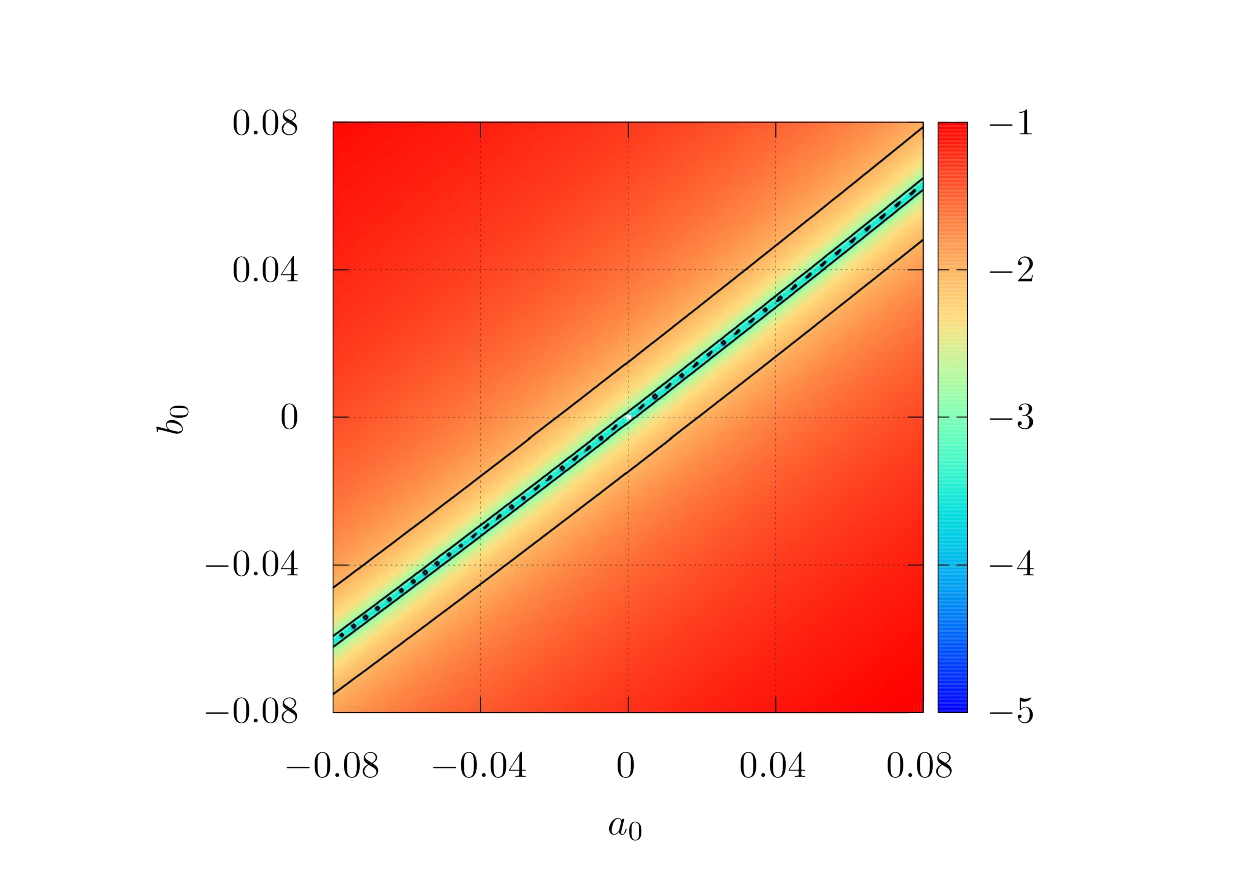}
	\hspace{-0.08\linewidth}
	\includegraphics[width=0.35\linewidth]{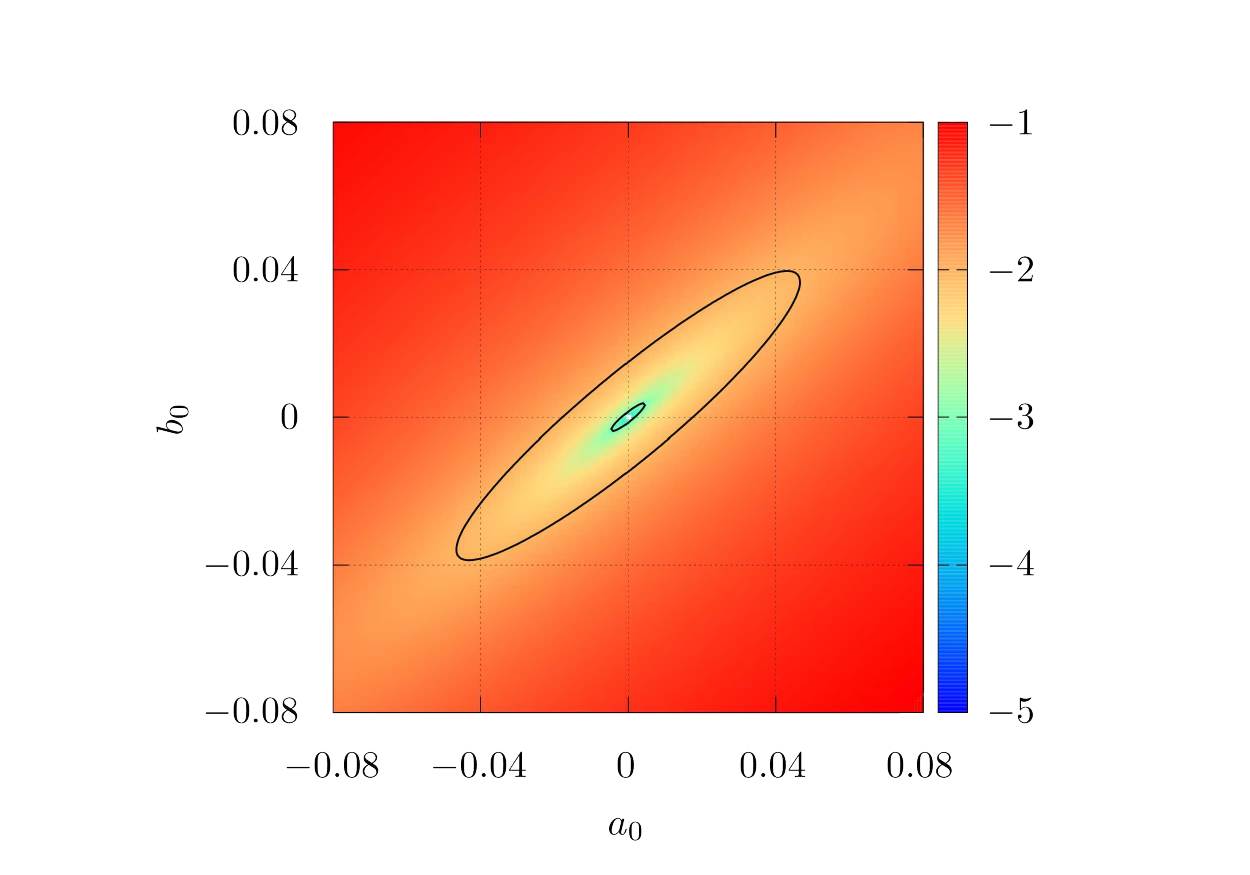}	
	\caption{Relative errors for $n=0$ and different $l$ for $[a_0, b_0]$. In the following we show the logarithm $\log_{10}(x)$ of the relative errors from left to right for: the real part $\omega_\text{r}$, the imaginary part $\omega_\text{i}$, and the combination $\omega_\text{c}$ eq. \eqref{combined_relative_error}. From top to bottom we show $l=2$ and $l=3$.\label{a0b0}}
\end{figure*}

\begin{figure*}[ht]
	\centering
	\includegraphics[width=0.35\linewidth]{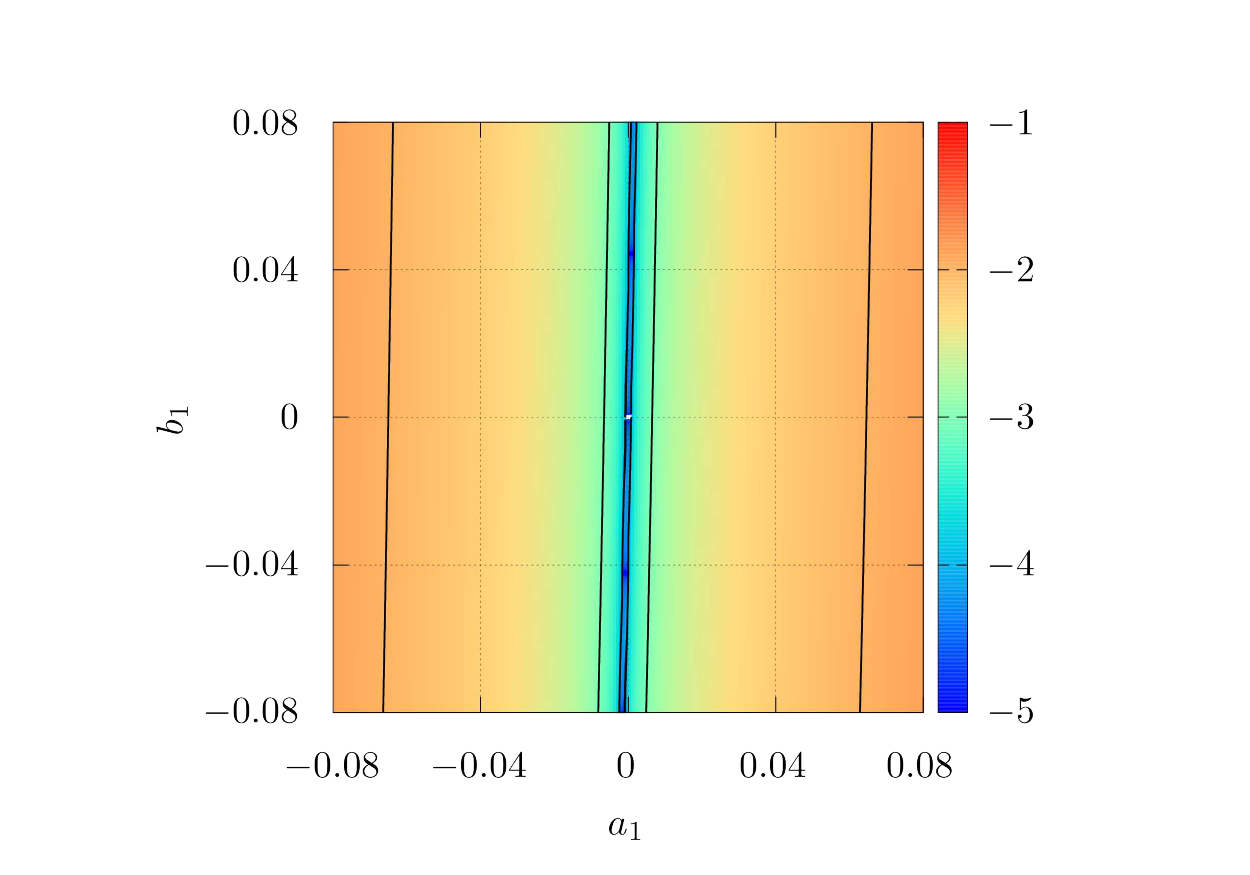}
	\hspace{-0.08\linewidth}
	\includegraphics[width=0.35\linewidth]{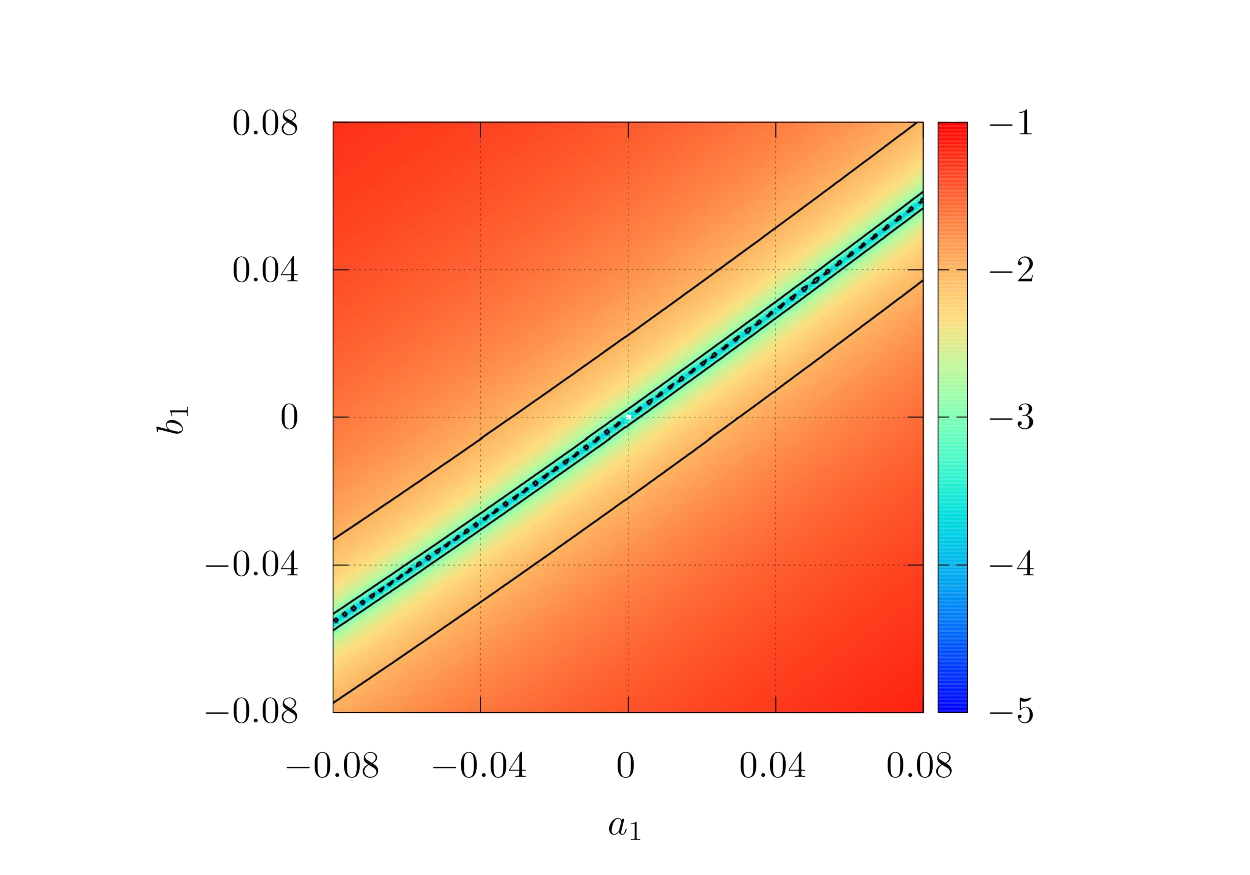}
	\hspace{-0.08\linewidth}
	\includegraphics[width=0.35\linewidth]{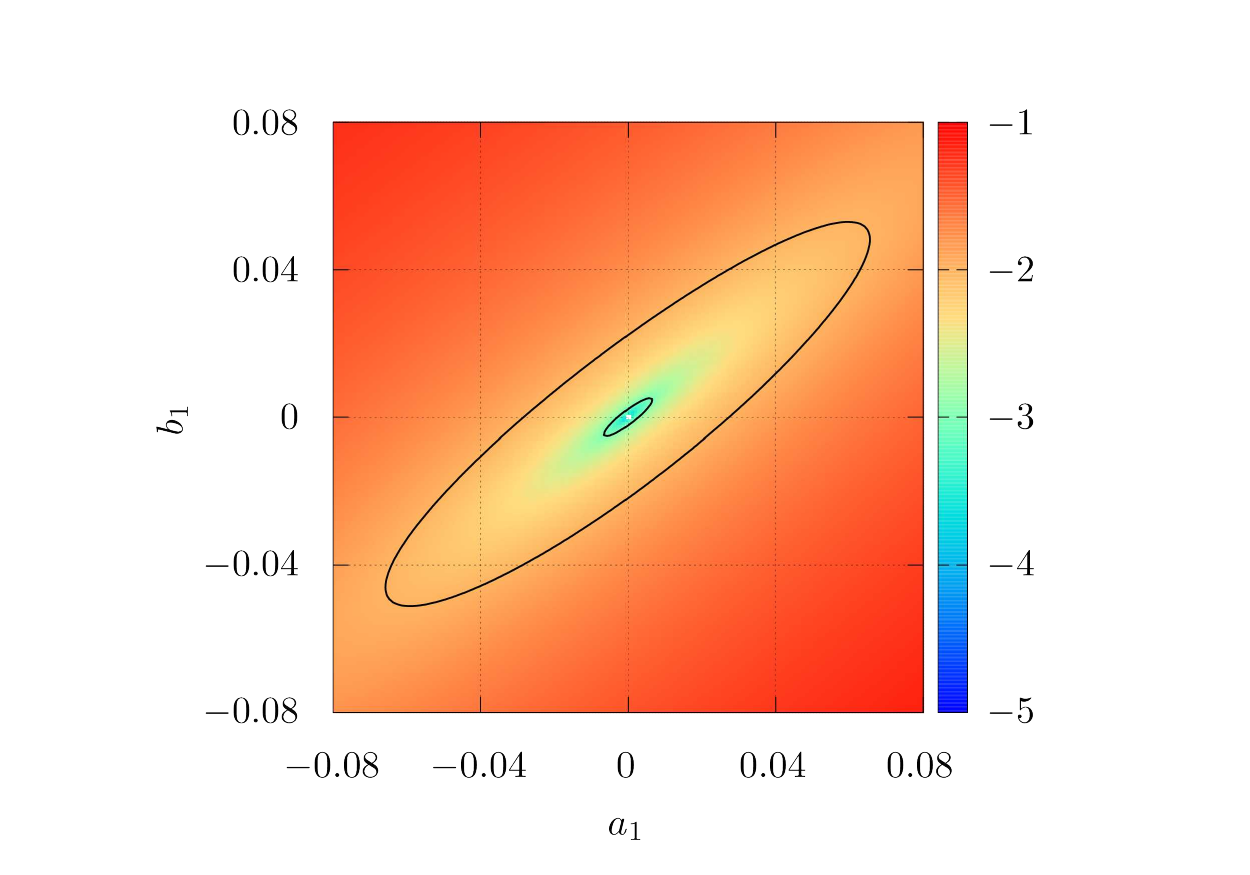}
	\\
	\centering
	\includegraphics[width=0.35\linewidth]{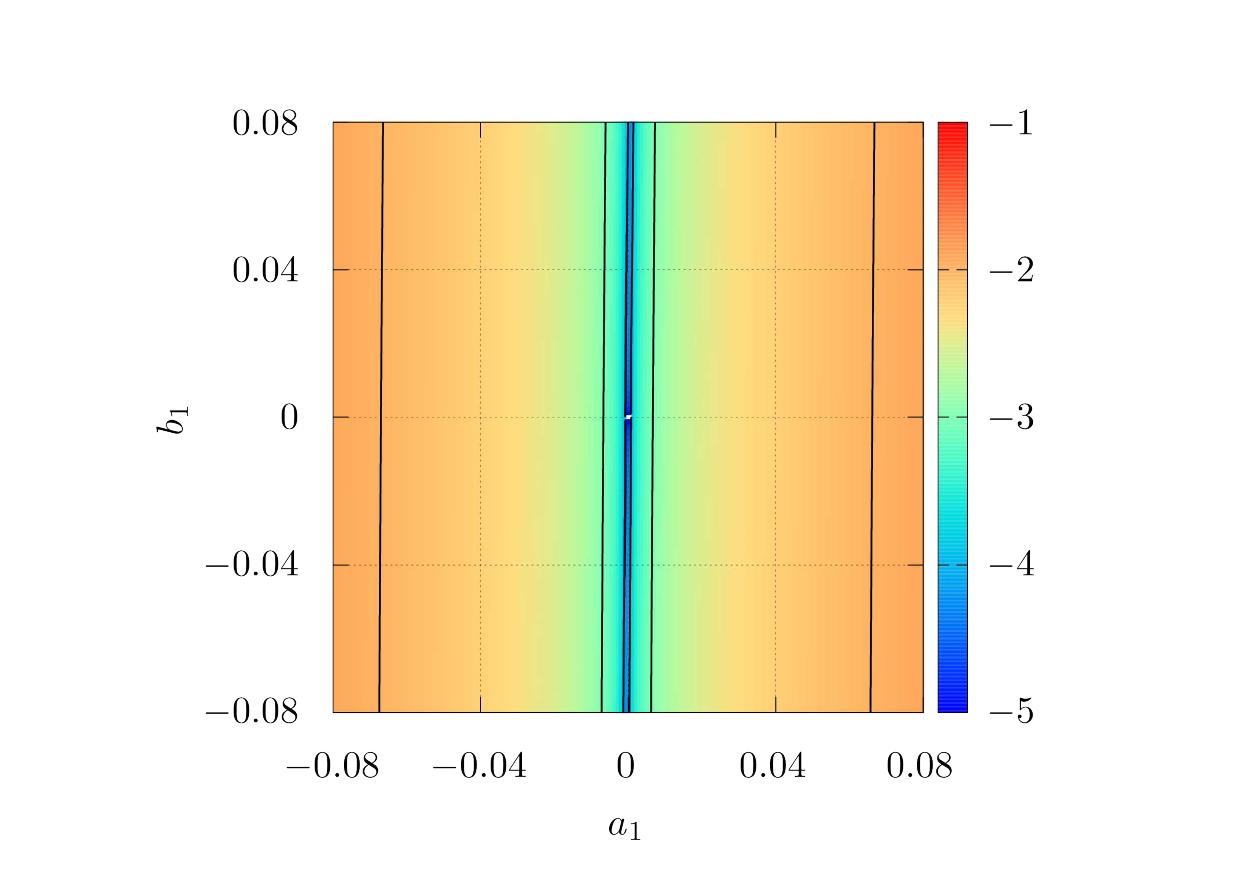}
	\hspace{-0.08\linewidth}
	\includegraphics[width=0.35\linewidth]{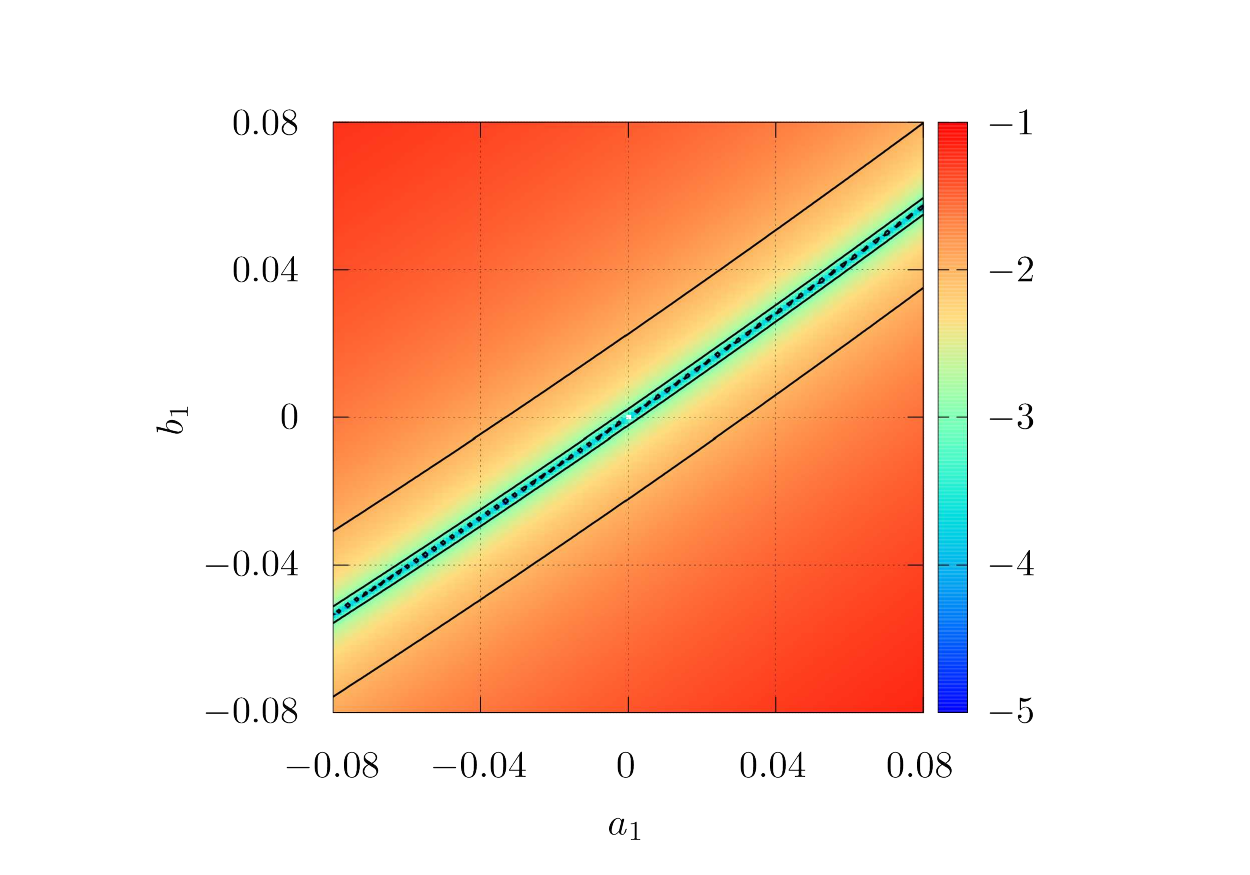}
	\hspace{-0.08\linewidth}
	\includegraphics[width=0.35\linewidth]{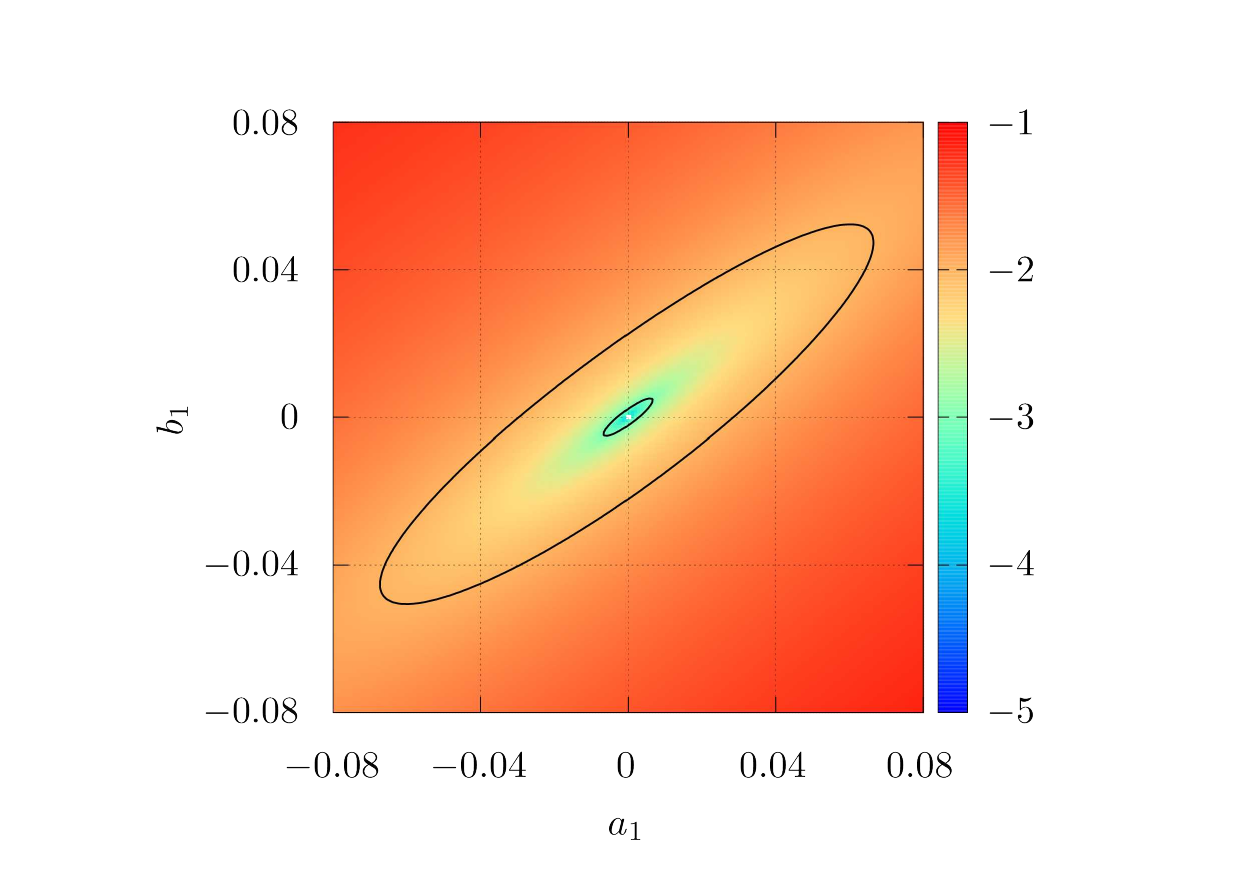}
	\caption{Relative errors for $n=0$ and different $l$ for $[a_1, b_1]$. In the following we show the logarithm $\log_{10}(x)$ of the relative errors from left to right for: the real part $\omega_\text{r}$, the imaginary part $\omega_\text{i}$, and the combination $\omega_\text{c}$ eq. \eqref{combined_relative_error}. From top to bottom we show $l=2$ and $l=3$.\label{a1b1}}
\end{figure*}

Here we show our results for the relative errors for the fundamental $(n=0)$ quasi-normal mode for $l=2$ (top panels) and $l=3$ (bottom panels) for $[a_0,b_0]$ in Fig. \ref{a0b0} and for $[a_1, b_1]$ in Fig \ref{a1b1}. The reference value to which the relative error is obtained is the $M=1$ Schwarzschild case for the same $n$ and $l$, computed with the WKB method. The relative errors in the real part $\omega_\text{r}$ are shown in the left panels and the ones for the imaginary part $\omega_\text{i}$ in the central panels. Finally, the combined relative errors $\omega_\text{c}$ are shown in the right panels. The color scaling shows the logarithm of the absolute value of the relative errors, respectively. The contour lines show integer values for this logarithm. 
\par
It is evident that the role of $[a_0, b_0]$ and $[a_1, b_1]$ is very similar for both $l$. Both can effectively influence each other in such a way that the real and imaginary parts remain unchanged compared to the Schwarzschild case. It is remarkable that the knowledge of only the real or imaginary parts for two different $l$ can hardly be used to constrain $[a_0, b_0]$ or $[a_1, b_1]$, because the scaling is very similar in both cases. It is by far more useful to connect the real and imaginary part of the same mode, which we show in the combined relative errors in the right panels. However, comparing this for both $l$, it is evident that merging both results is of limited usefulness in the more precise determination of the parameters. Also, note that the real part is almost blind to changes in $b_0$ and $b_1$, while the imaginary part is sensitive to both.
\subsection{Parameters $[M,a_0,b_0]$}\label{results-2}
\begin{figure*}[ht]
	\includegraphics[width=0.45 \linewidth]{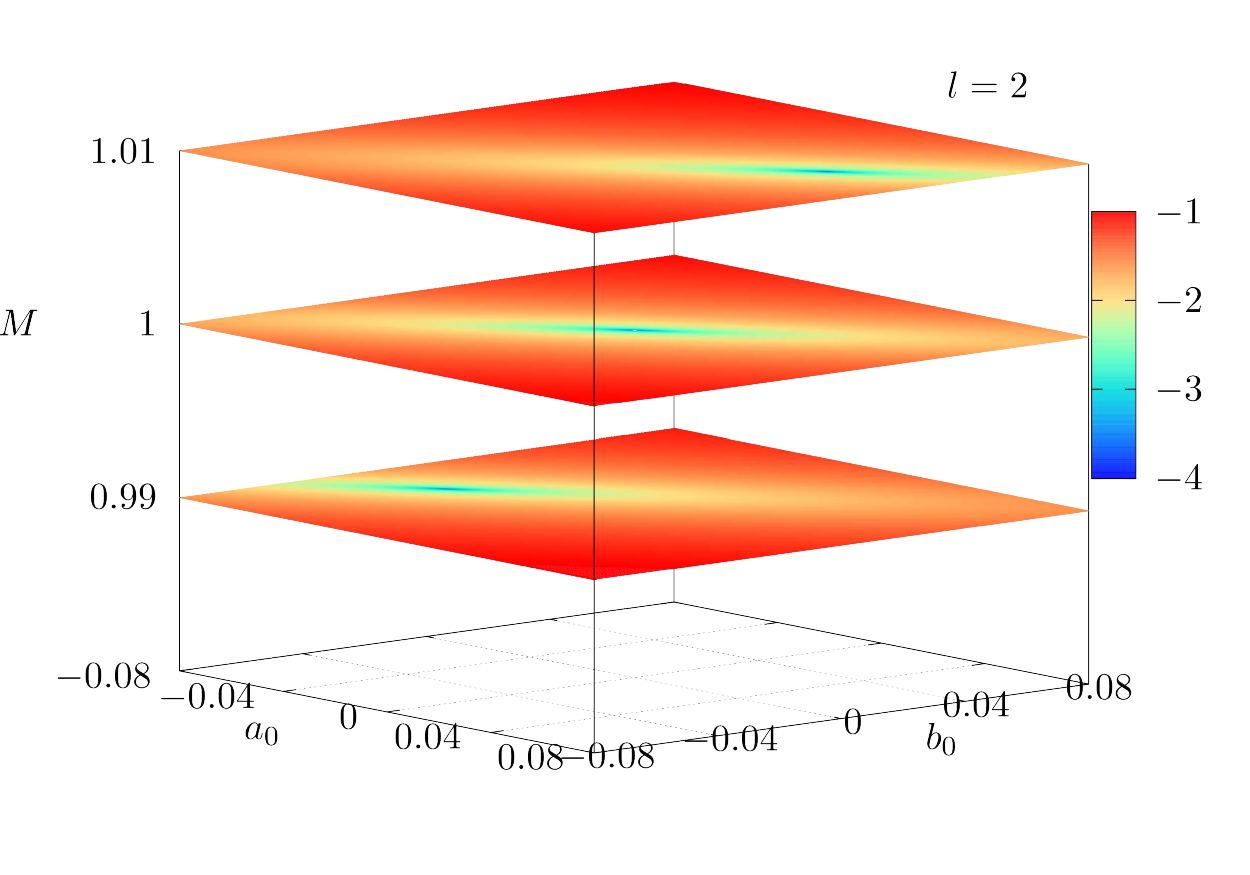}
	\includegraphics[width=0.45 \linewidth]{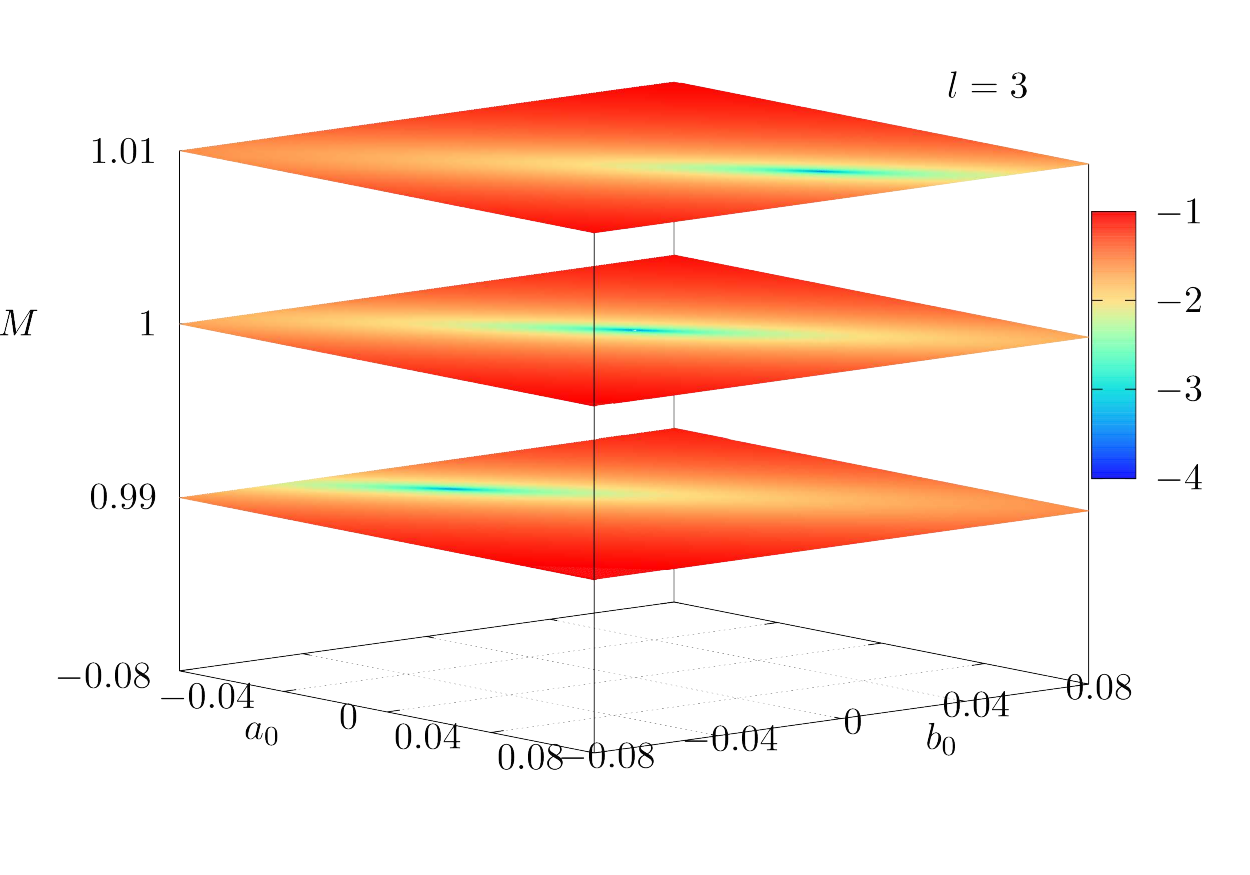}
	\\
	\includegraphics[width=0.45 \linewidth]{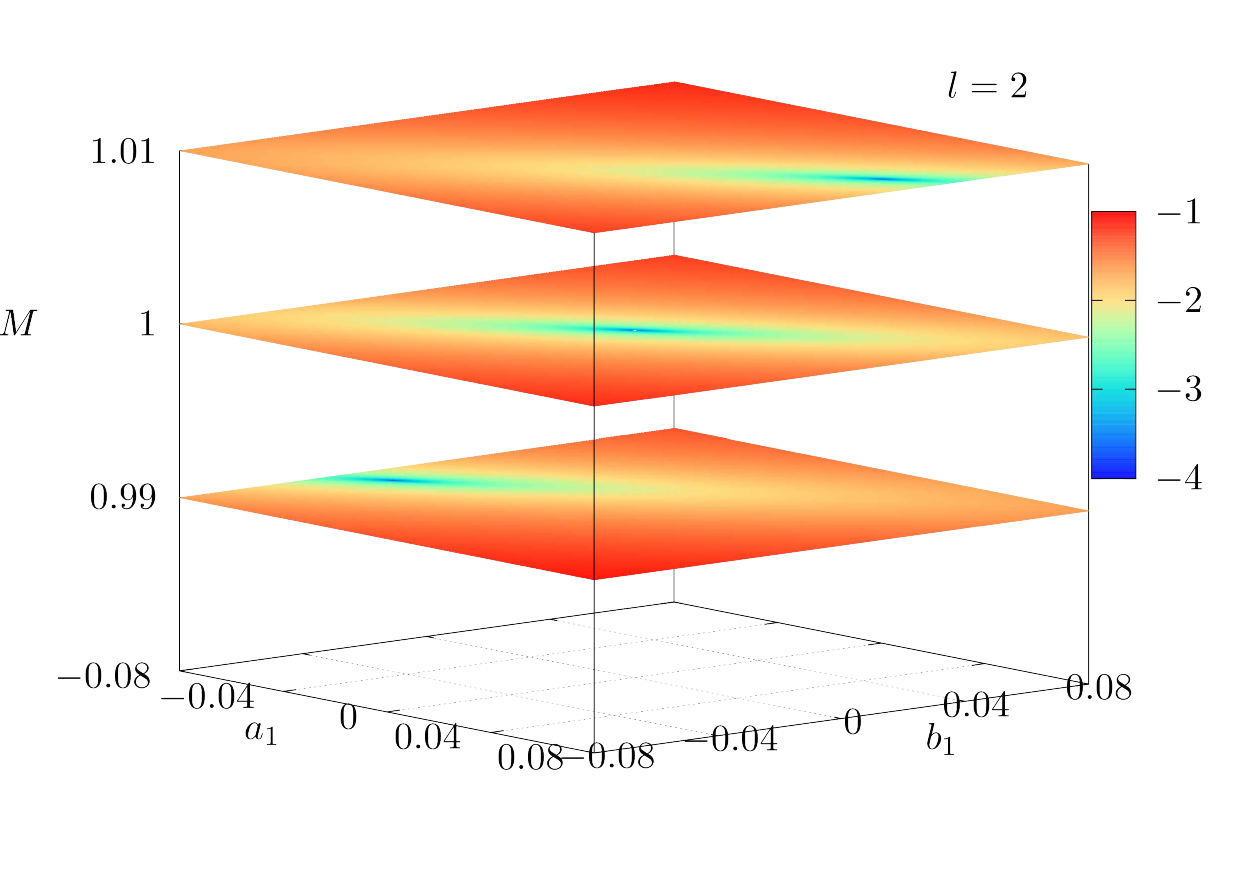}
	\includegraphics[width=0.45 \linewidth]{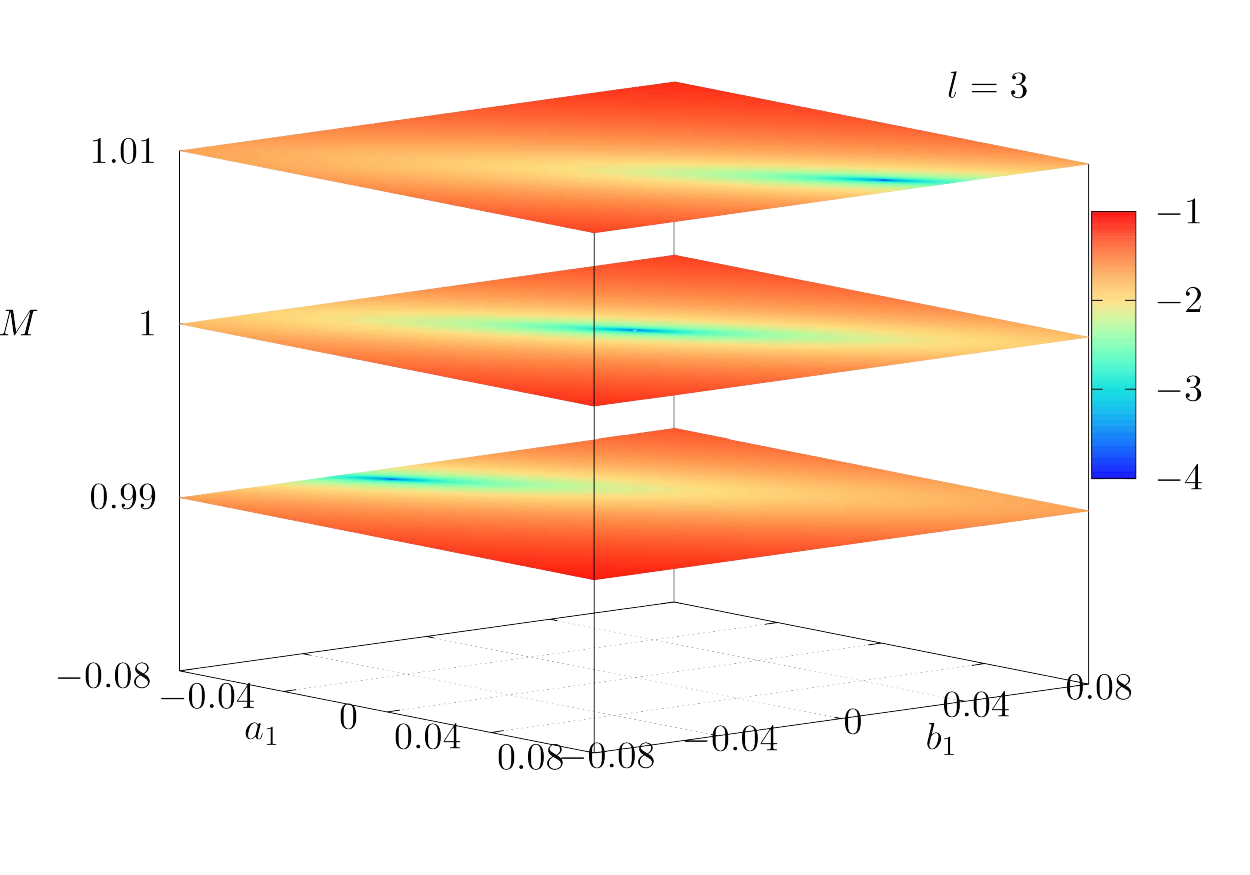}
	\caption{Here we show the combined relative errors $\omega_\text{c}$ for $n=0$ and different $M$. The top panels are the $l=2$ (left) and $l=3$ (right) case for $[a_0, b_0]$, while the same $l$ cases are shown in the bottom panels for $[a_1,b_1]$. Each layer is obtained for different mass $M$, while setting the non-varying parameters to the general relativity value of $0$. \label{surface}}
\end{figure*}

In this section we investigate up to which extent a change in the mass $M$ can allow the fundamental quasi-normal modes of non-vanishing $a_0$ and $b_0$ to mimic the pure Schwarzschild modes with reference mass $M=1$. Since it is more involved to visualize the contours of the now three-dimensional parameter space, we decided to show slices of constant $M$ in the $a_0$ and $b_0$ parameter plane. These two dimensional slices through the three dimensional parameter space are orthogonal to the $M$-axis and have a similar color map representation as used in the previous Sec. \ref{results-1}. This is shown in the top panels of Fig. \ref{surface}.
\par
The similar scaling of the combined relative errors for $l=2$ and $l=3$ is remarkable. It shows that changing the mass by $1\,\%$ allows values for $[a_0, b_0]$ of around $\pm0.04$, assuming the relative error would be $10^{-2}$. Since the scaling for both $l$ is very similar, merging the two parameter ranges can hardly be used to constrain the mass. The relative errors for the real and imaginary parts scale also very similar, which we do not show explicitly here.
\subsection{Parameters $[M,a_1,b_1]$}\label{results-3}
In the bottom panels of Fig. \ref{surface} we show our results for repeating the analysis from Sec. \ref{results-2} with varying $a_1$ and $b_1$. Now $a_0$ and $b_0$ are set to zero. Qualitatively we arrive at a very similar result. Again it is possible to find non-zero values for the parameters through modifying the black hole mass. But, since the combined error areas enclosed by the same contour are larger here, compared to the $[a_0, b_0]$ case, the combination $[a_1, b_1]$ requires higher precision to be constrained.
\subsection{Investigating the First Overtone $n=1$}\label{results-4}
\begin{figure*}[ht]
	\centering
	\includegraphics[width=0.35\linewidth]{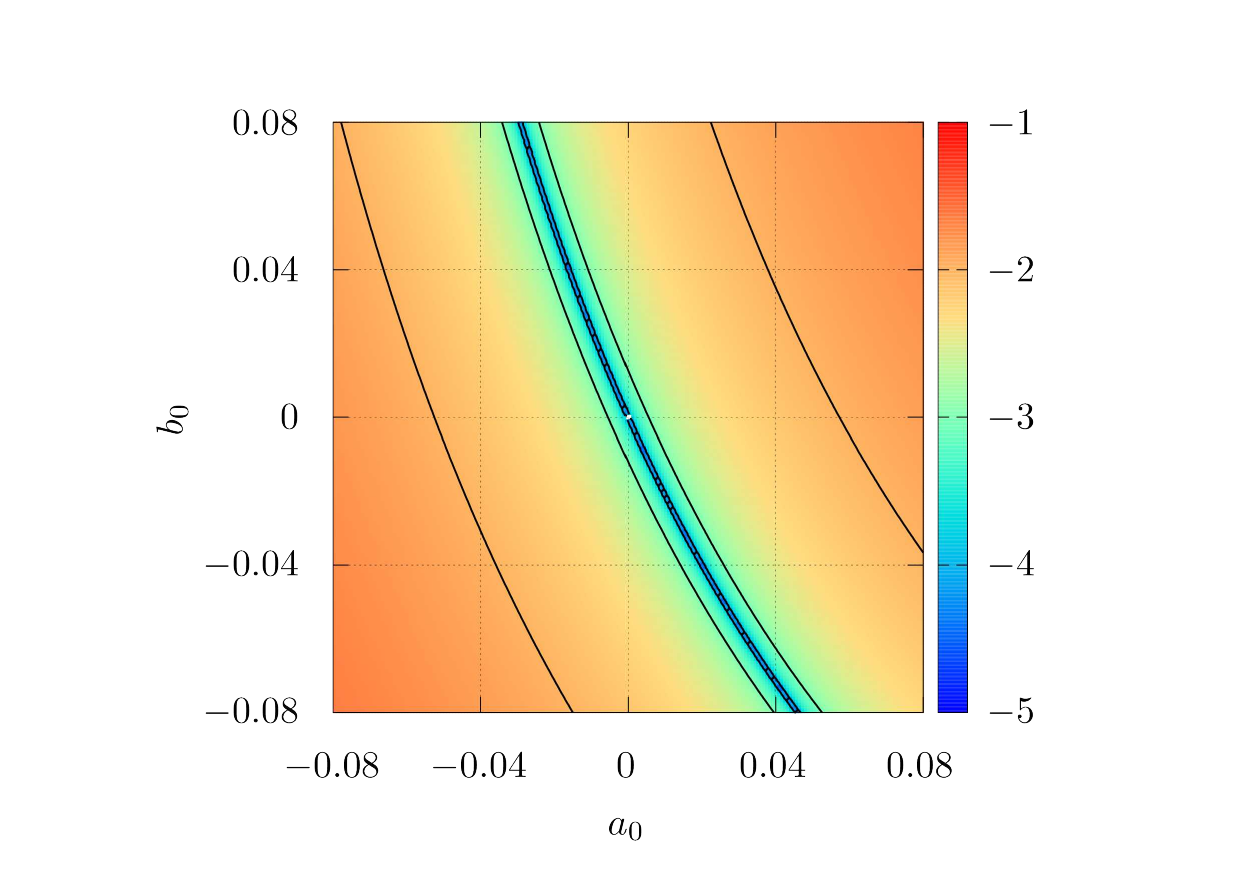}
	\hspace{-0.08\linewidth}
	\includegraphics[width=0.35\linewidth]{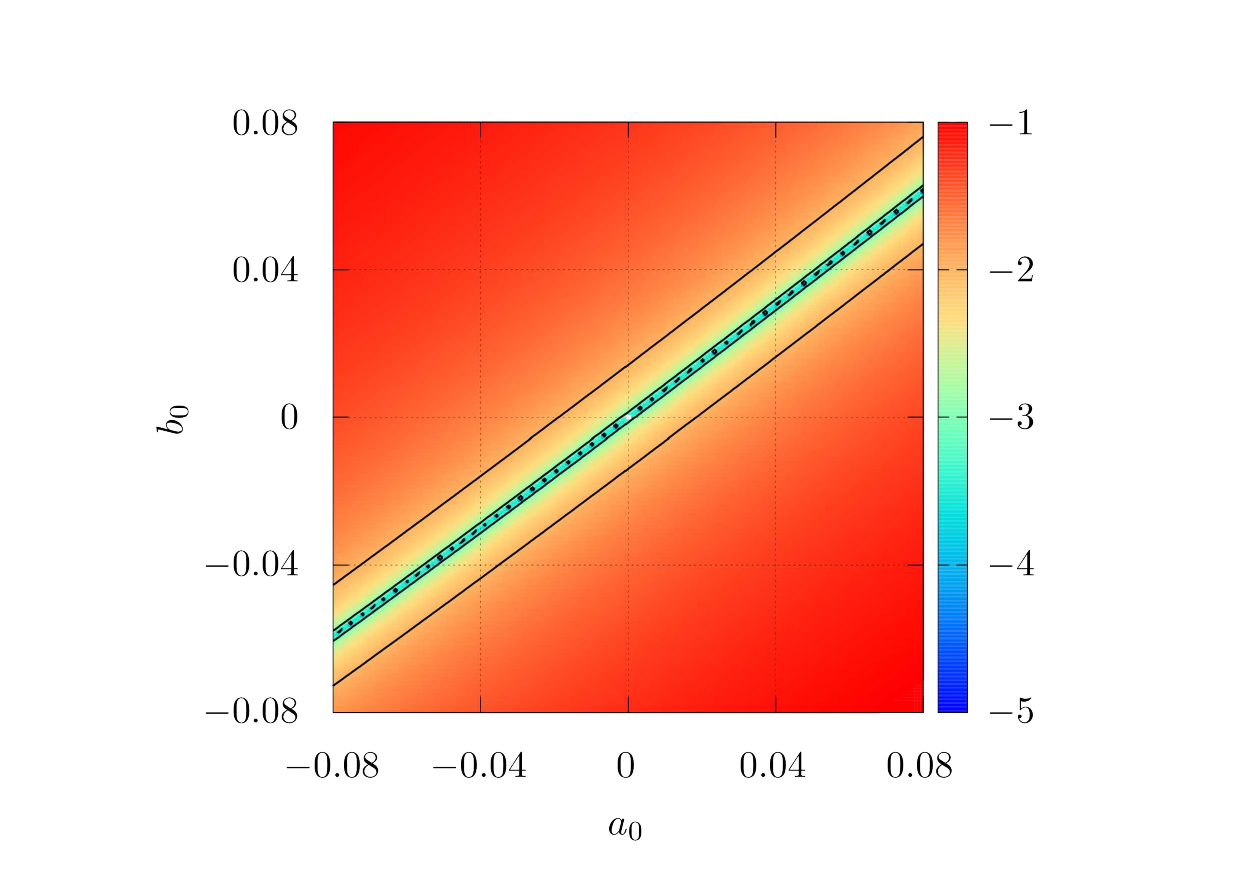}
	\hspace{-0.08\linewidth}
	\includegraphics[width=0.35\linewidth]{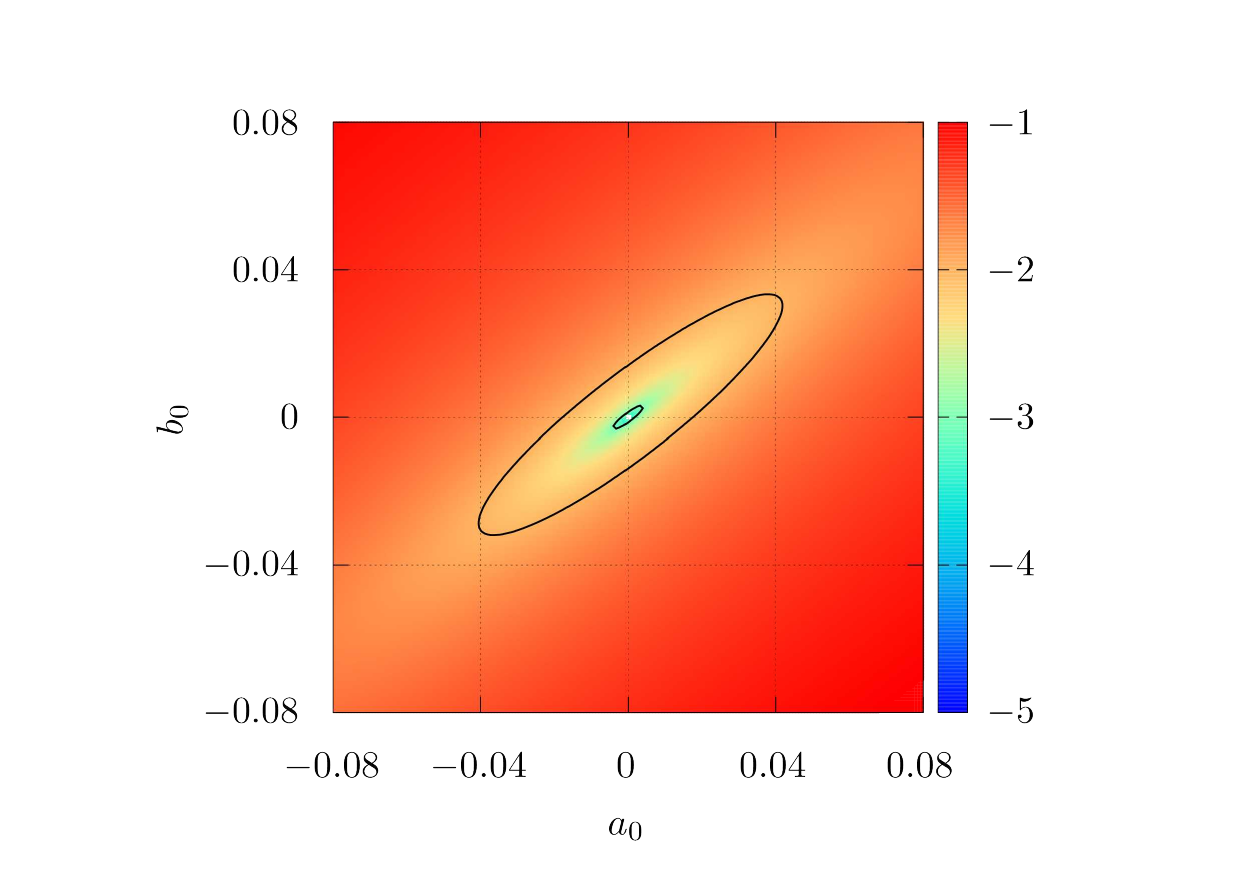}
	\\
	\centering
	\includegraphics[width=0.35\linewidth]{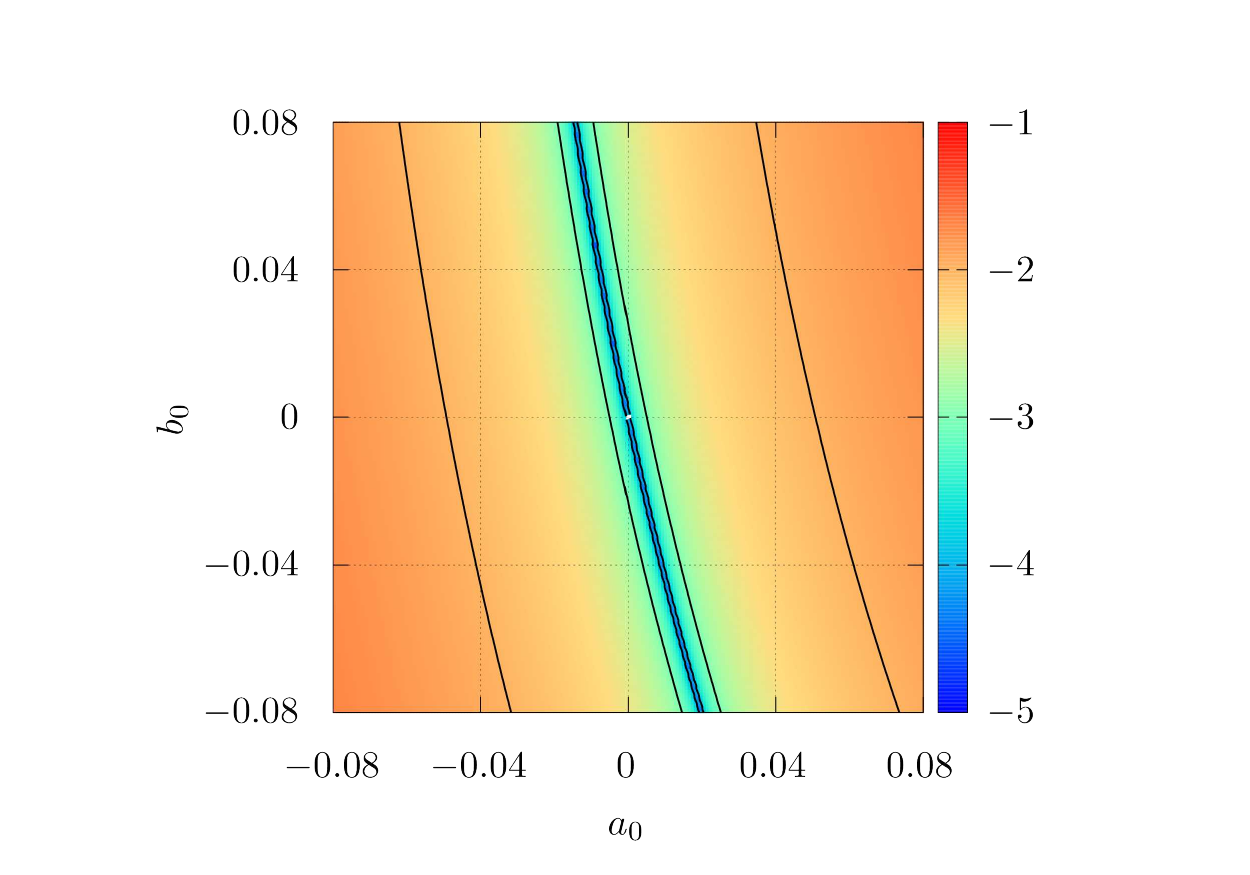}
	\hspace{-0.08\linewidth}
	\includegraphics[width=0.35\linewidth]{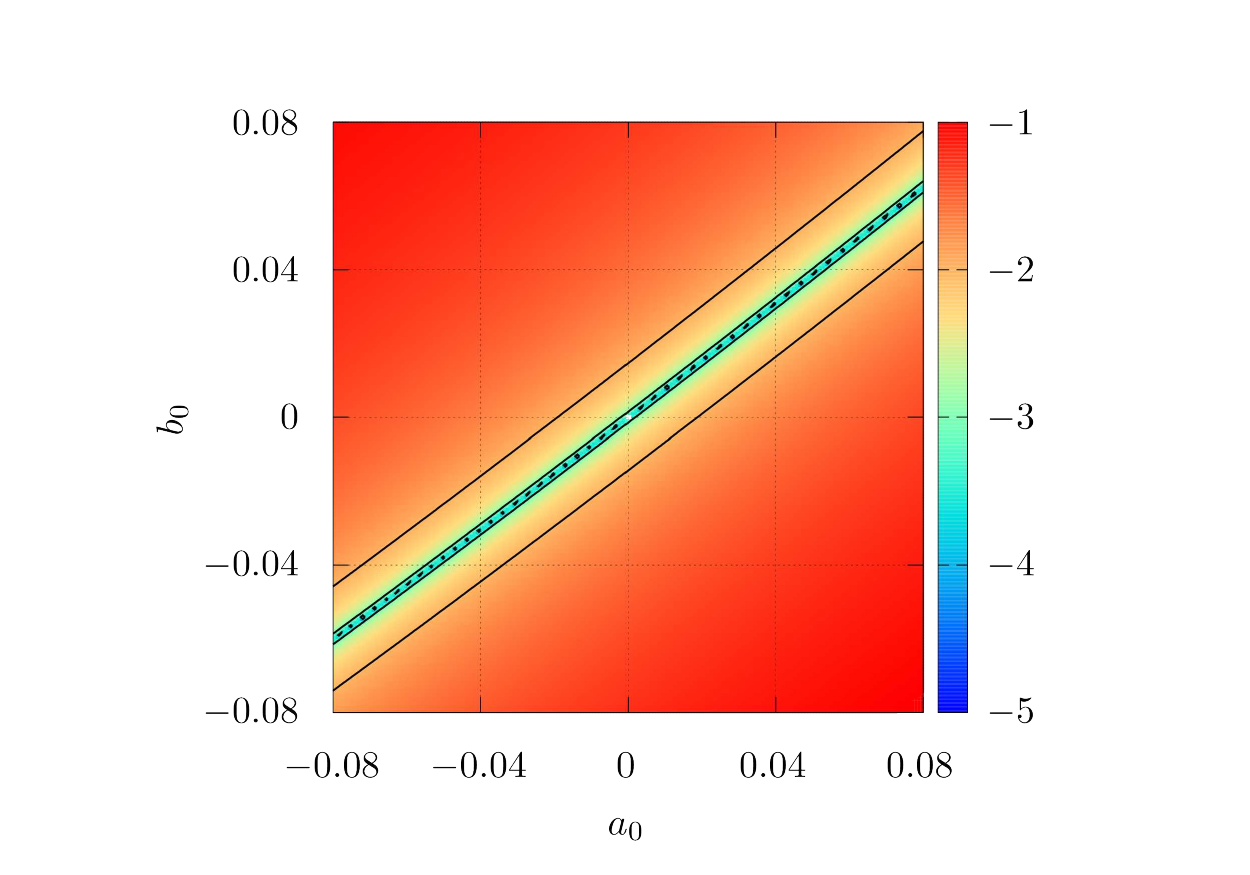}
	\hspace{-0.08\linewidth}
	\includegraphics[width=0.35\linewidth]{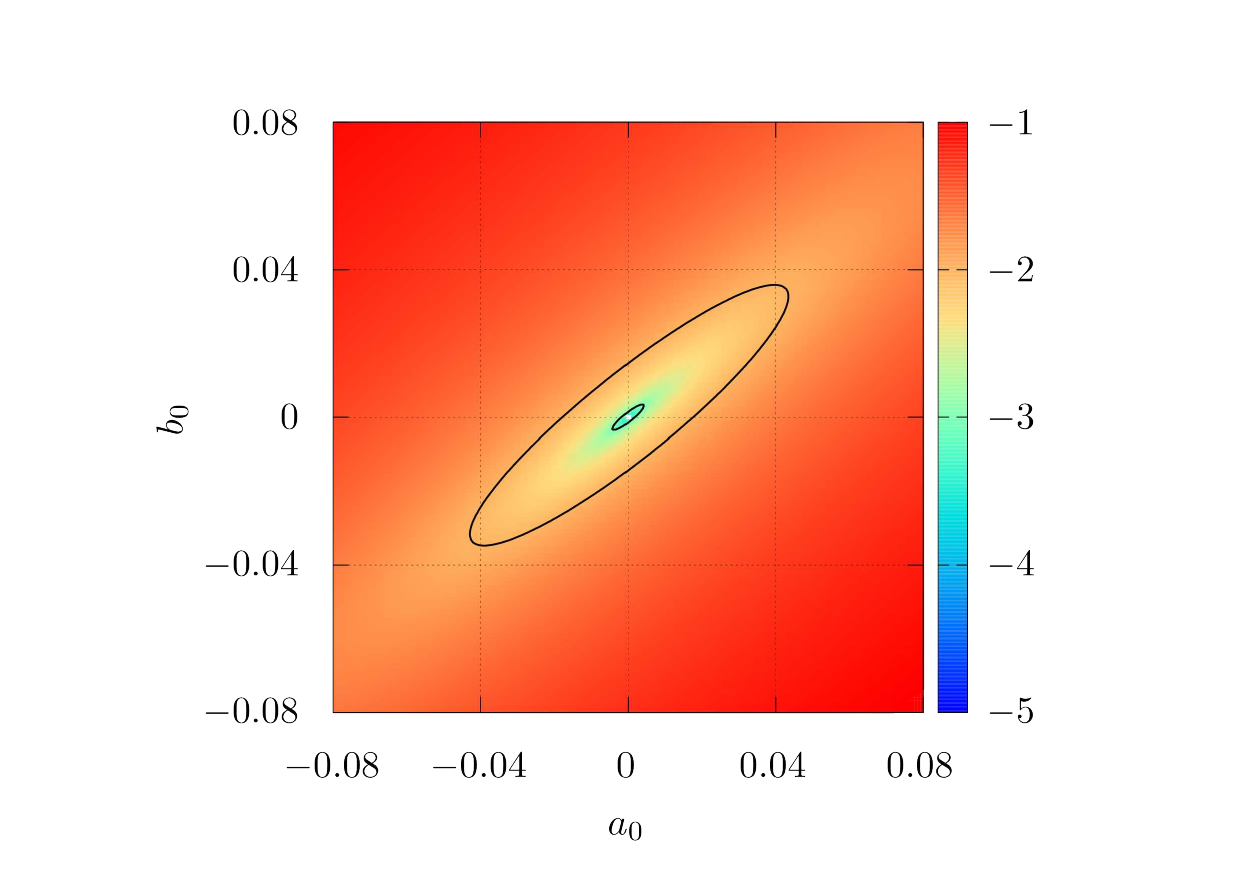}	
	\caption{Relative errors for $n=1$ and different $l$ for $[a_0, b_0]$. In the following we show the logarithm $\log_{10}(x)$ of the relative errors from left to right for: the real part $\omega_\text{r}$, the imaginary part $\omega_\text{i}$, and the combination $\omega_\text{c}$ eq. \eqref{combined_relative_error}. From top to bottom we show $l=2$ and $l=3$.\label{a0b0_n1}}
\end{figure*}

\begin{figure*}[ht]
	\centering
	\includegraphics[width=0.35\linewidth]{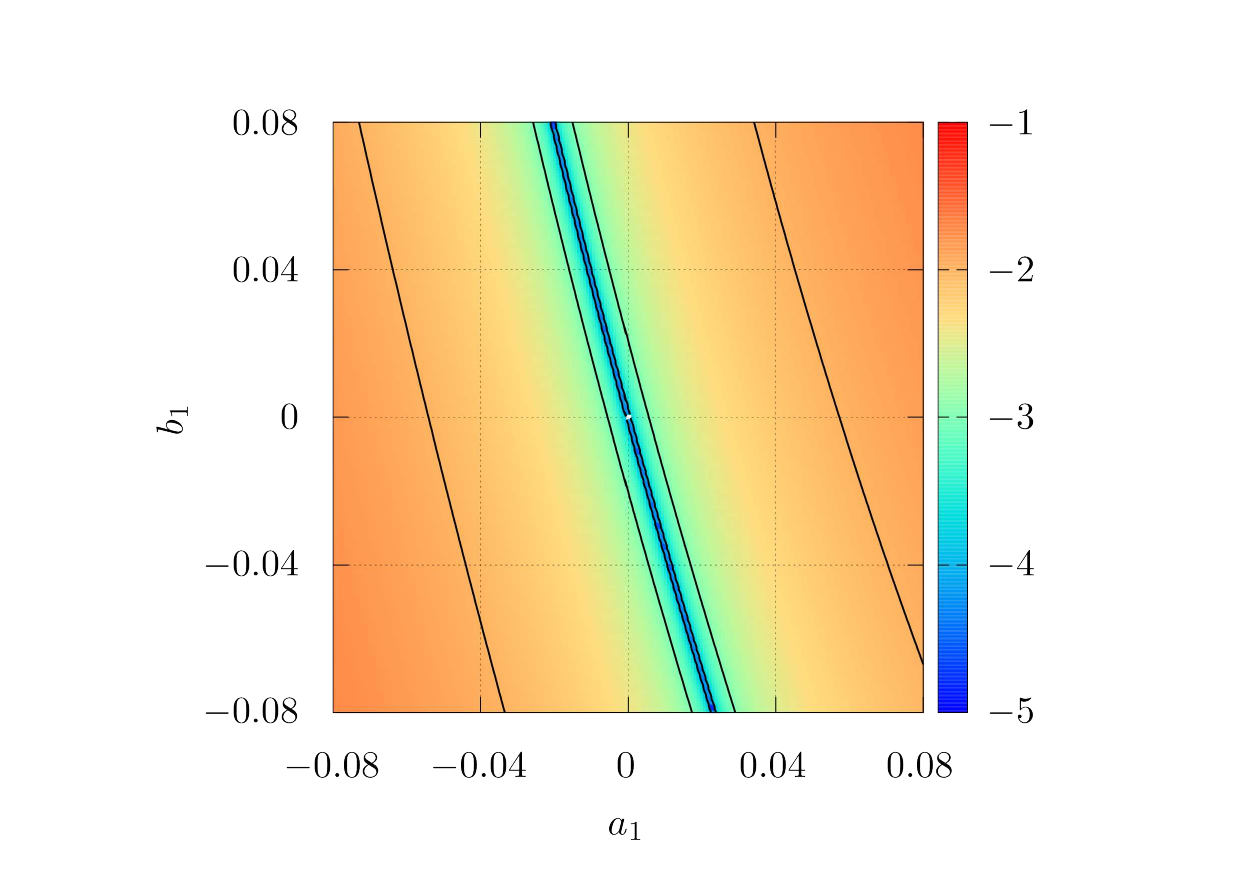}
	\hspace{-0.08\linewidth}
	\includegraphics[width=0.35\linewidth]{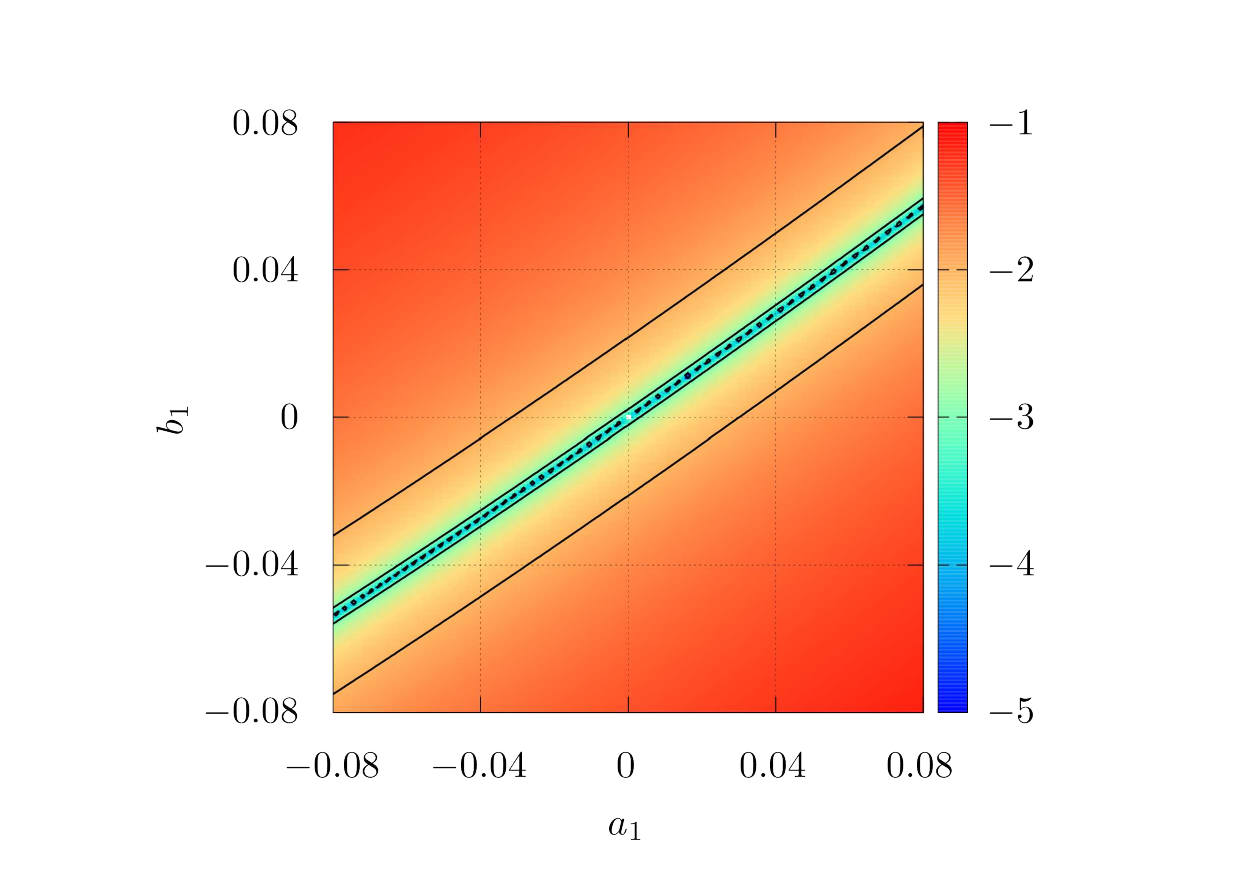}
	\hspace{-0.08\linewidth}
	\includegraphics[width=0.35\linewidth]{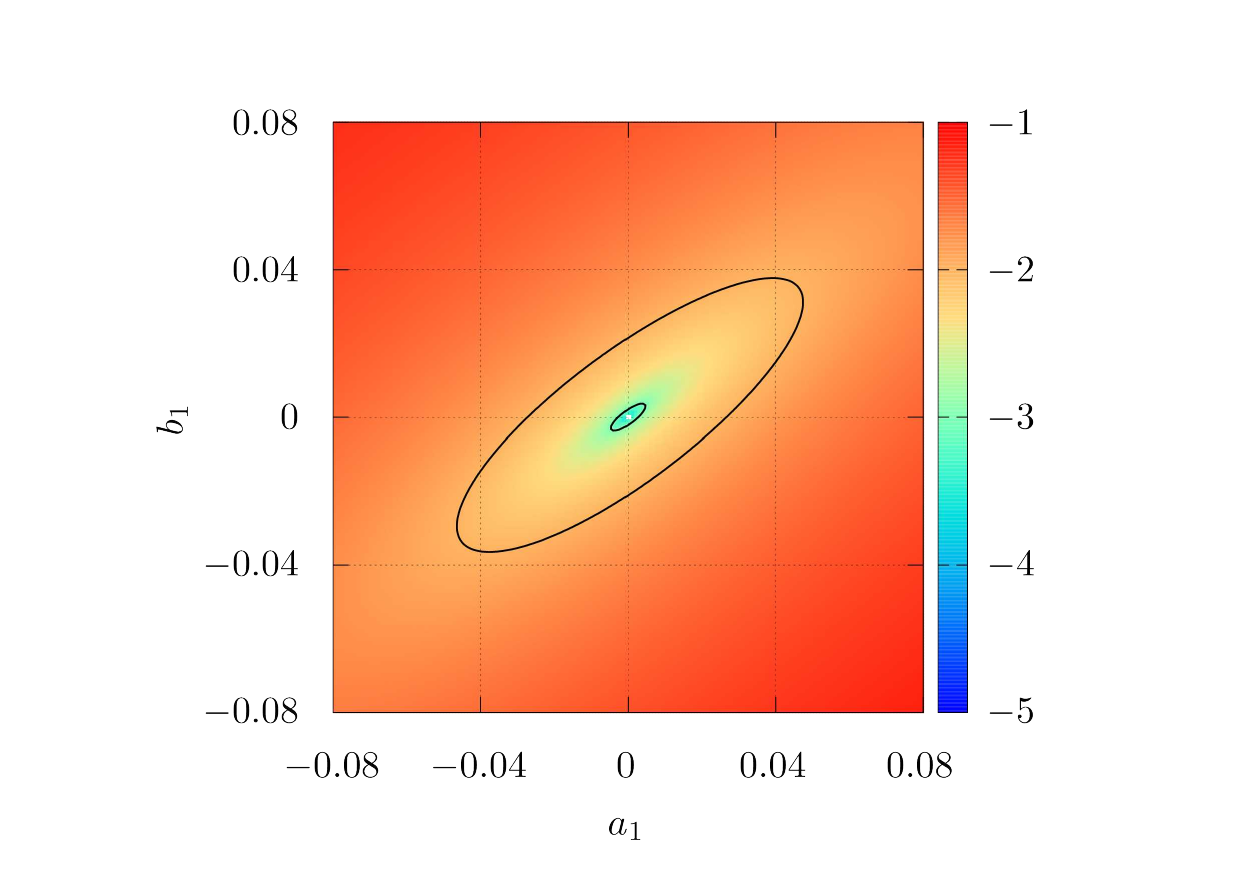}
	\\
	\centering
	\includegraphics[width=0.35\linewidth]{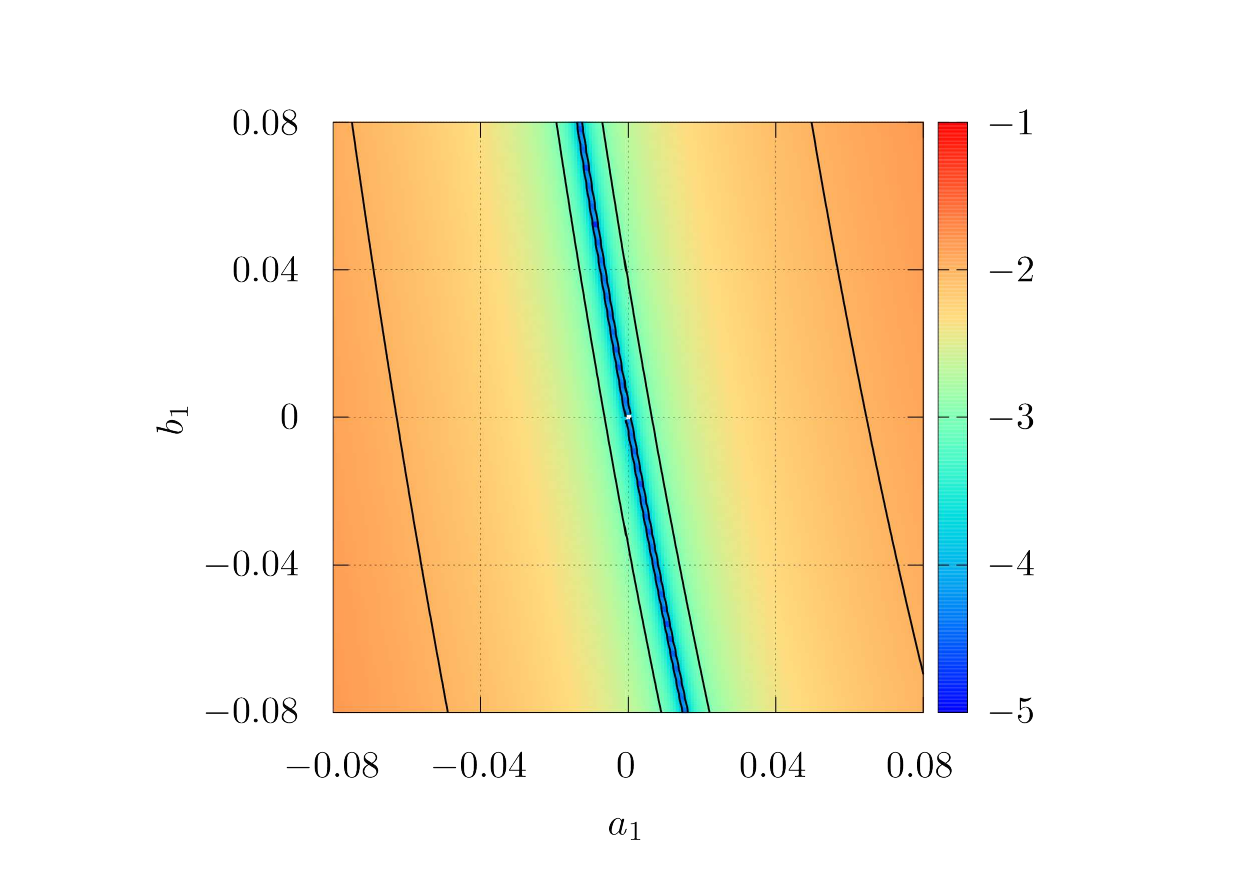}
	\hspace{-0.08\linewidth}
	\includegraphics[width=0.35\linewidth]{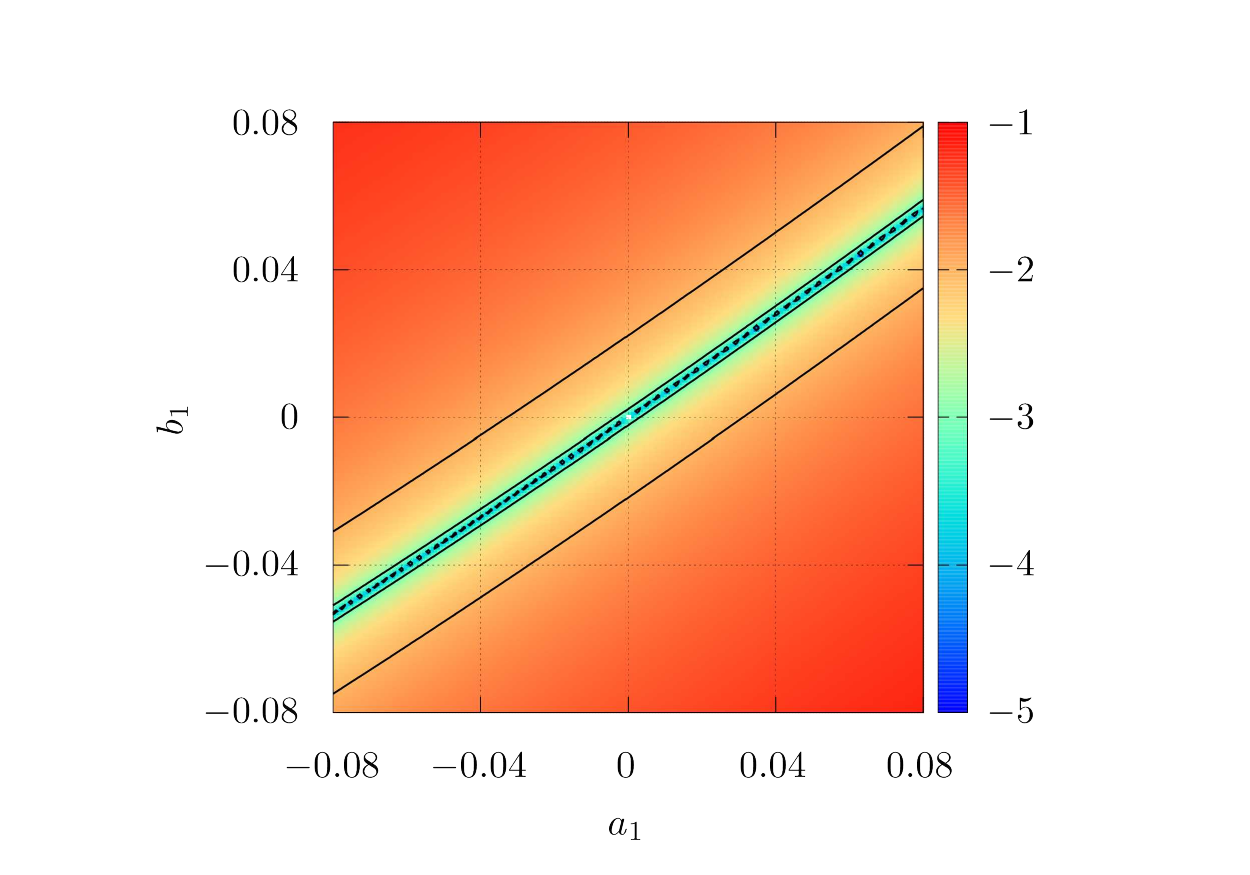}
	\hspace{-0.08\linewidth}
	\includegraphics[width=0.35\linewidth]{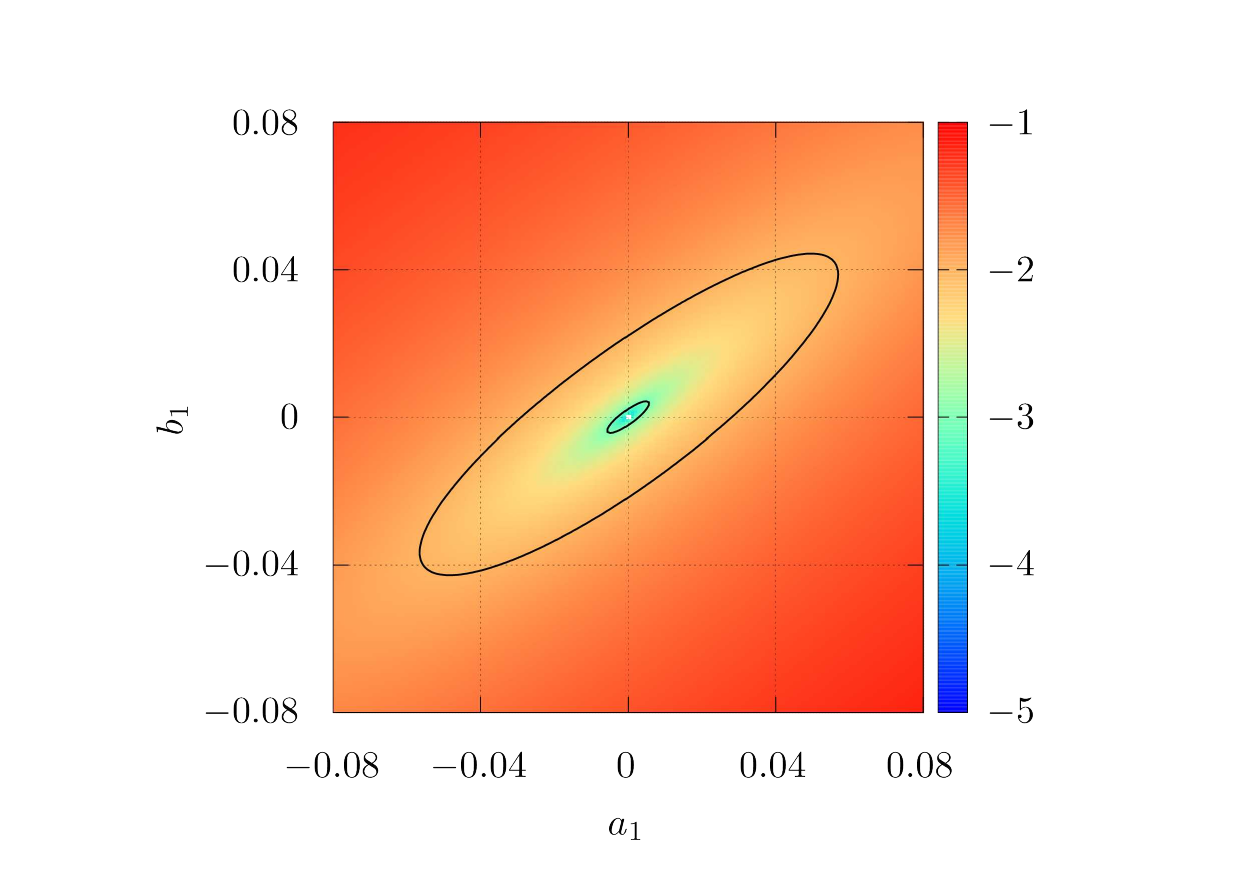}
	\caption{Relative errors for $n=1$ and different $l$ for $[a_1, b_1]$. In the following we show the logarithm $\log_{10}(x)$ of the relative errors from left to right for: the real part $\omega_\text{r}$, the imaginary part $\omega_\text{i}$, and the combination $\omega_\text{c}$ eq. \eqref{combined_relative_error}. From top to bottom we show $l=2$ and $l=3$.	\label{a1b1n1}}
\end{figure*}

\begin{figure*}[ht]
	\includegraphics[width=0.45 \linewidth]{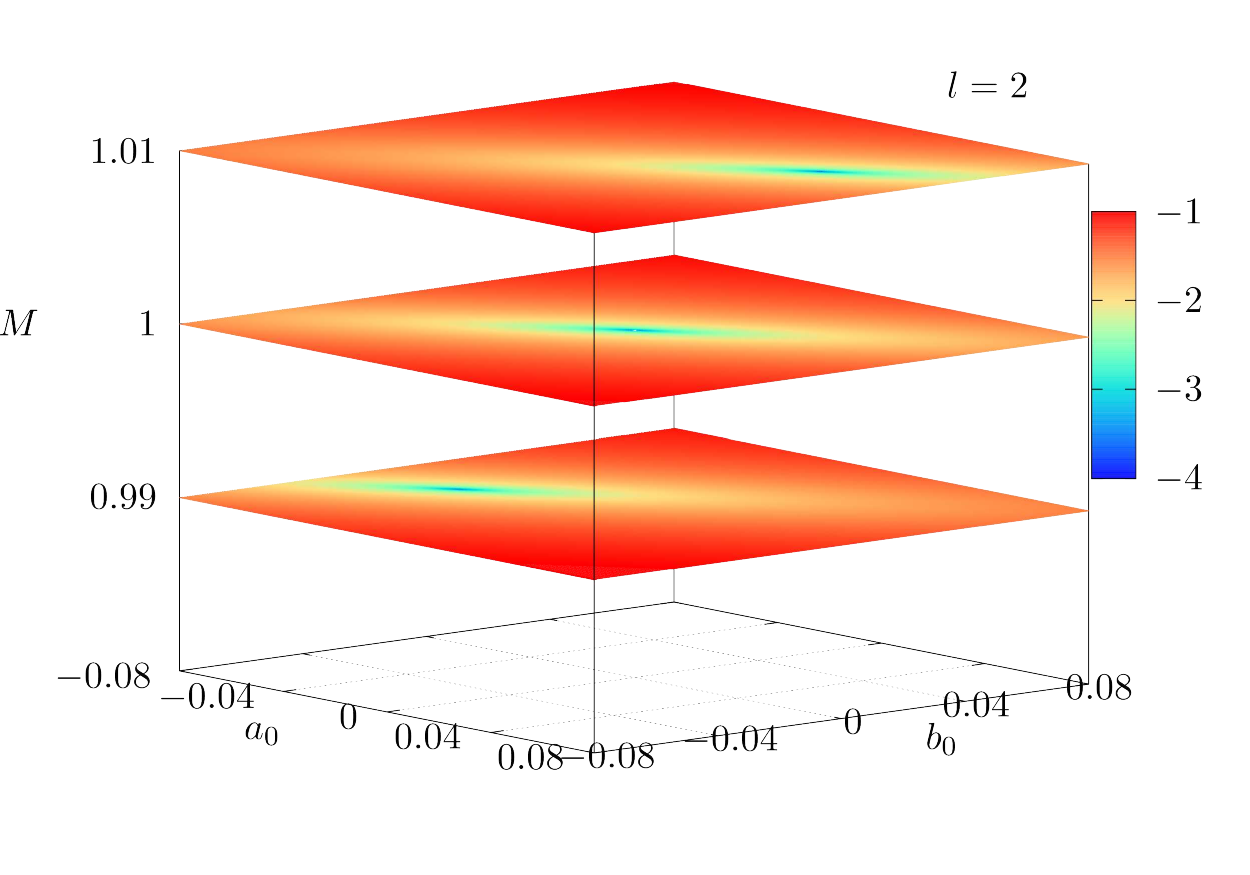}
	\includegraphics[width=0.45 \linewidth]{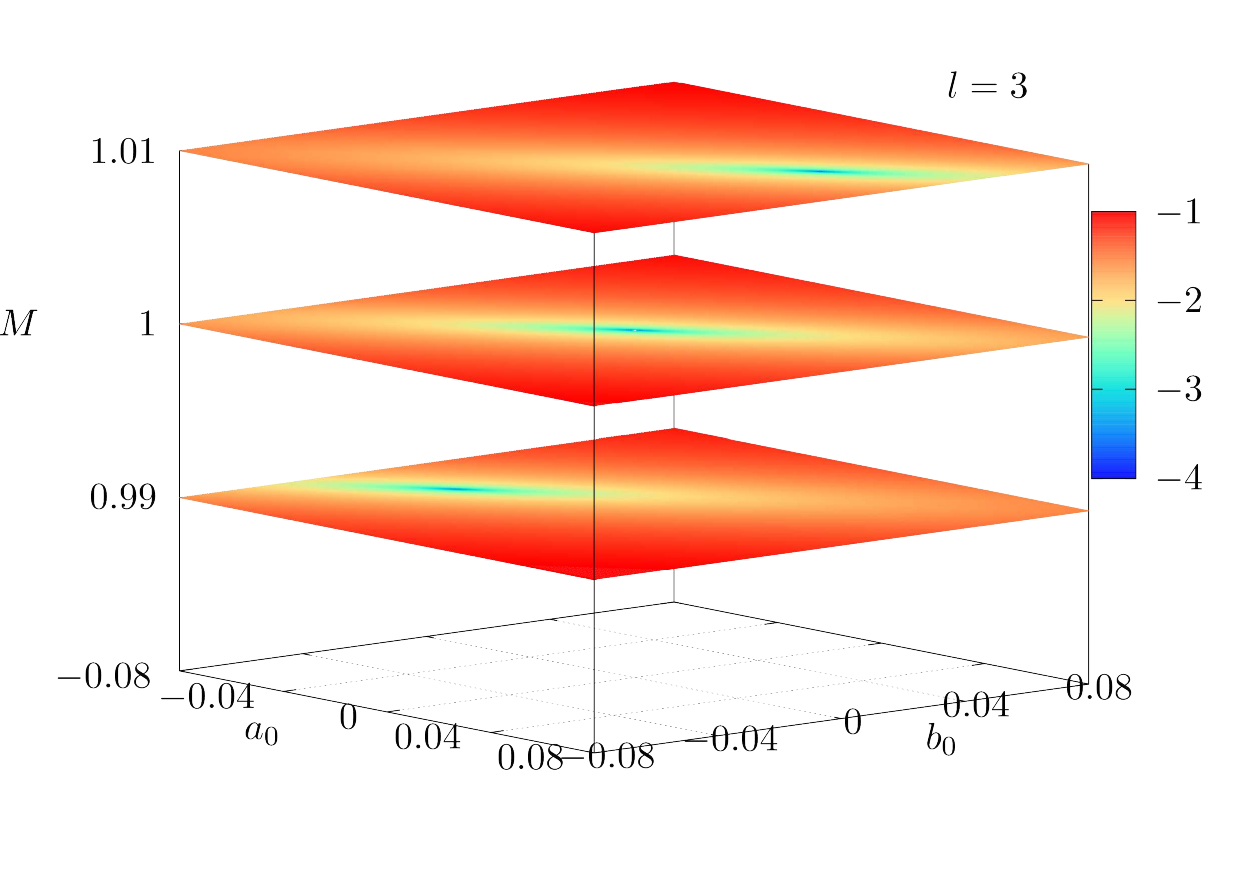}
	\\
	\includegraphics[width=0.45 \linewidth]{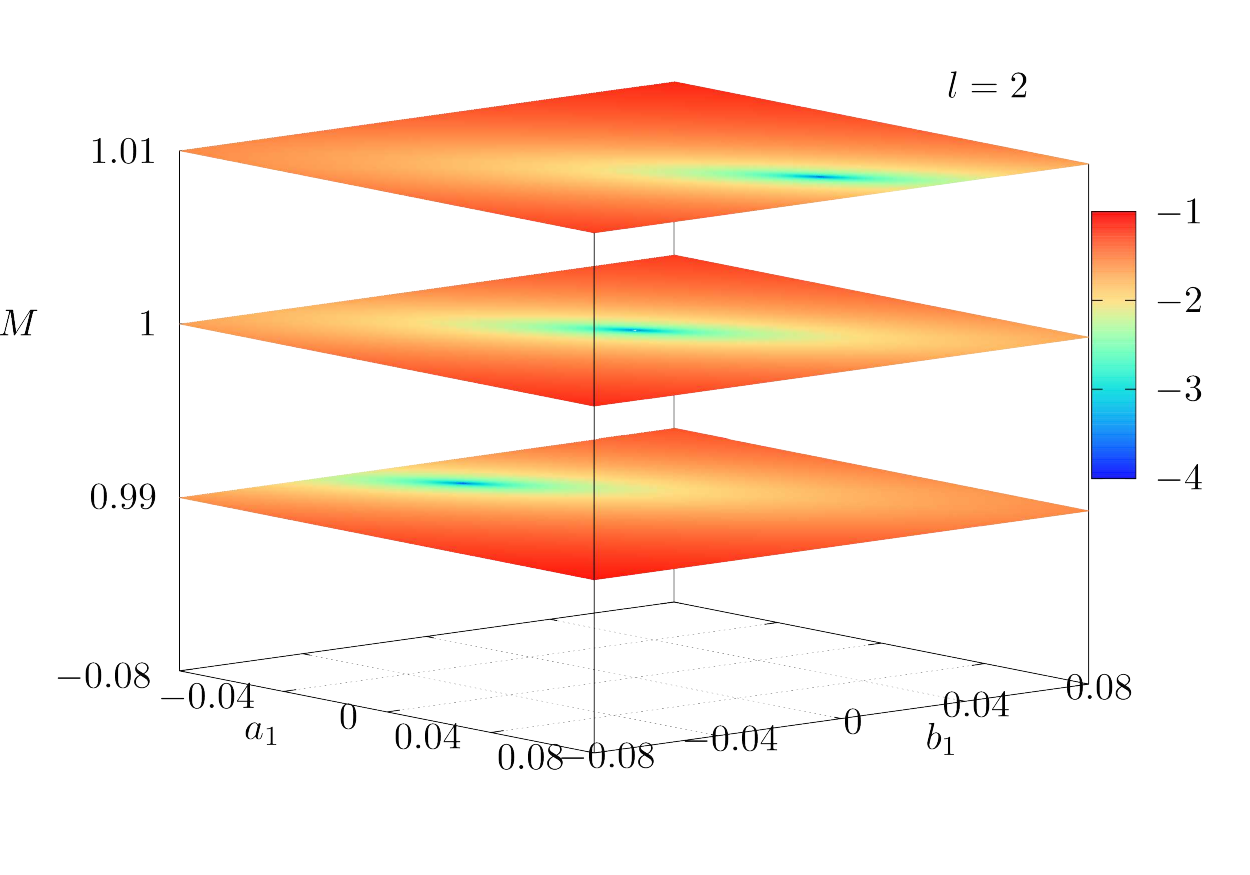}
	\includegraphics[width=0.45 \linewidth]{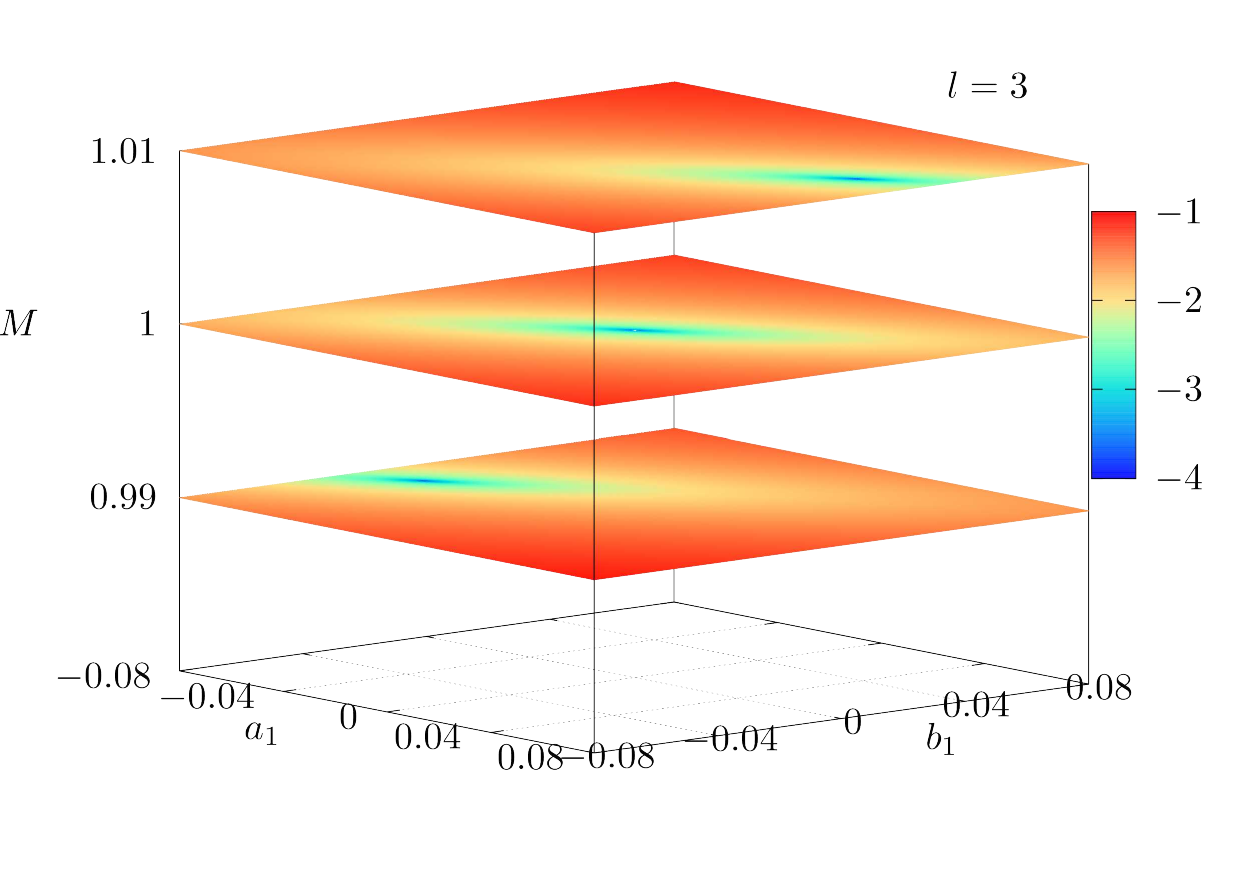}
	\caption{Here we show the combined relative errors $\omega_\text{c}$ for $n=1$ and different $M$. The top panels are the $l=2$ (left) and $l=3$ (right) case for $[a_0, b_0]$, while the same $l$ cases are shown in the bottom panels for $[a_1,b_1]$. Each layer is obtained for different mass $M$, while setting the non-varying parameters to the general relativity value of $0$. \label{surface_n1}}
\end{figure*}

The analysis of the previous subsections can also be done for the first overtone $n=1$, which we show in Fig. \ref{a0b0_n1}, Fig. \ref{a1b1n1}, and Fig. \ref{surface_n1}. We find that the overtone changes the scaling of the real part (left panels), it is now more sensitive to changes in $b_0$ and $b_1$. The change in the imaginary part (central panels) seems to be more stable towards the first overtone $n$. However, unless pristine precision is available, the combined error contours (right panels) overlap significantly in all studied cases and can hardly be used to constrain the parameters $[a_0, b_0]$ to be within their PPN limits or $[a_1,b_1]$ to be of similar order.
\subsection{The Effective Potentials}\label{effective potentials}
In order to verify that changing the mass parameter $M$ can actually recast the Schwarzschild fundamental quasi-normal modes, although $[a_0, b_0]$ and $[a_1, b_1]$ are non-zero, we show specific effective potential barriers in Fig. \ref{potential}. These serve as verification that the application of the WKB formula is well justified, because the potential barriers have two turning points and very similar behavior to the well established Schwarzschild case.
\par
From all panels one can deduce that changing the mass by $1\,\%$ around the reference value of $M=1$ leads to clearly distinguishable potential barriers for Schwarzschild (solid lines). This is expected and will change the quasi-normal modes in the same order. In all panels, we also show multiple realizations of the RZ metric (dashed lines), where we choose $[a_0, b_0]$ and $[a_1, b_1]$ to be the ones that minimize the combined errors $\delta \omega_\text{c}$ (here for $n=0$). Note that these values slightly differ from $l=2$ to $l=3$ and are not exactly the same. However, since the contours are almost the same for both $l$, the values are very close to each other, as shown in Fig. \ref{surface}. 
\par
For large $r^*$, all potentials go to zero, as expected from construction of the metric. Around the peak of the barrier, which is the region of the potential that determines the properties of the fundamental quasi-normal modes, the RZ potentials match the Schwarzschild reference potential extremely well, although their mass parameter is different. At the same time, the other Schwarzschild potentials differ the most. Going to large negative values of $r^*$, which means to approach the black hole horizon, we see that the dashed lines in the $[a_1,b_1]$ case deviate slightly more than the ones for $[a_0,b_0]$. This should be expected, since non-zero values of $[a_1,b_1]$ become by construction more significant closer to the horizon, but less important for large $r^*$.
\par
Finally, the precise match of the constructed RZ potentials with the Schwarzschild reference potential also confirms that the WKB method is well suited for the problem.
\section{Discussion}\label{discussion}
We first discuss our findings for varying $[a_0,b_0]$ and $[a_1,b_1]$ simultaneously for reference mass $M=1$. Afterwards, we interpret the effect of allowing different values for the mass for these combinations. The reconstructed potentials, as well as the role of rotation are discussed in the end.
\par
From our findings in Sec. \ref{results-1} we see that knowing only either the real or imaginary parts of the fundamental quasi-normal modes for $l=2$ and $l=3$ is less useful than knowing real and imaginary part of the same $l$. This is because combinations of $[a_0,b_0]$ and $[a_1,b_1]$ leaving either the real or imaginary part of the Schwarzschild result unchanged, behave quite similar for both $l$. Combining the real and imaginary parts is more powerful when constraining the parameters. Looking into the absolute numbers, we find that one requires a very precise knowledge of the real and imaginary parts of a given quasi-normal mode in order to constrain $[a_0,b_0]$ to the already known limits coming from the PPN expansion, which is discussed in Sec. \ref{parametrization of the metric} and was derived in \cite{PhysRevD.90.084009}. Note that this section assumes that the mass is provided without any error.
\par
However, treating the mass $M$ as free parameter in Sec. \ref{results-2}, we report combinations in Fig. \ref{surface}, such that one obtains a similar result for the fundamental quasi-normal modes of different $l$. In fact, introducing an uncertainty in the mass $M$ of around one percent allows values for $a_0$ and $b_0$ to be of order $10^{-2}$, which is beyond the PPN constraints of $\sim 10^{-4}$, \cite{PhysRevD.90.084009}. However, realizing such high precision in determining $[a_0,b_0]$ requires pristine knowledge of $M$. Achieving this in actual observations is challenging and might require combined measurements using a network of third generation gravitational wave detectors \cite{Maselli:2017kvl}. Since the scaling of the relative errors for both $l$ is quite similar, the combination of both fundamental modes is not of significant help. In practice, this means that the mass parameter $M$ is clearly degenerate with respect to some combinations of $a_0$ and $b_0$. We conclude that even small uncertainties in $M$ can actually mimic Schwarzschild fundamental quasi-normal modes for suitable choices of $[a_0,b_0]$.
\par
In Sec. \ref{results-3} we arrived at a very similar results for combinations $[a_1,b_1]$. Note that these parameters are not connected in the same way to the PPN parameters and should, by construction of the RZ metric, be more significant close to the horizon than far away. We find a similar scaling with respect to $M$. Quantitatively the error contours are a bit larger compared to $[a_0,b_0]$, making these parameters eventually more difficult to be constrained by quasi-normal modes.
\par
We also studied the role of the first overtone $n=1$ in Sec. \ref{results-4}. There we find that it can be helpful in determining parameters of the metric, but the contours still follow qualitatively similar patterns. Again, without very high precision for the provided spectrum, determining the leading parameters $[a_0,b_0]$ within their PPN limits seems unlikely. However, especially for introducing changes in the scaling of the real part, the access to overtones looks very promising. The role of overtones in testing the no-hair theorem and doing parameter estimation, are ongoing research \cite{Baibhav:2018rfk,Breschi:2019wki,Giesler:2019uxc}.
\par
With the explicit demonstration of the effective potentials in Sec. \ref{effective potentials}, we have verified that the shape of the potential allows the application of the third order WKB formula. This also motivates that our findings are reasonable, because the reconstructed potential barriers agree excellently. But, since the WKB method is not an exact method, small deviations with respect to exact methods are possible. Finally, because differences between the RZ potentials and the reference Schwarzschild potential become only visible close to the horizon, it is reasonable to expect that time-evolution calculations describing scattered radiation should give almost indistinguishable results too.
\par
Now we want to make some remarks regarding the possible application of the RZ metric in the inverse spectrum problem, where one reconstructs the perturbation potentials or metric from a given quasi-normal mode spectrum, e.g. \cite{paper2,paper5,Suvorov:2018bvs,paper6,Konoplya:2018ala}. An important question in this problem is the uniqueness of the reconstructed potential. In some of the mentioned works, Birkhoff's theorem \cite{Jebsen2005,Birkhoff} for spherically symmetric and non-rotating space-times was used. However, in this work, we do not have such a theorem if the underlying theory is not known. At the same time, it is well understood that different potentials can admit the same spectrum. Popular examples are the perturbation potentials of axial and polar perturbations described by the Regge-Wheeler \cite{1957PhRv..108.1063R} and Zerilli potentials \cite{1970PhRvL..24..737Z}. This was shown by Chandrasekhar and Detweiler \cite{Chandrasekhar441} and connected to the Darboux transform recently \cite{PhysRevD.96.024036}. Considering only the first few parameters in the RZ metric, and demanding them to match Schwarzschild like fundamental modes, will give tight constraints on the potential around the maximum. Since the RZ parameters introduce, by construction, more significant effects close to the horizon, than far away, this puts additional constraints. Note that the potentials constructed in this work actually match the Schwarzschild potential in these regions extremely well. However, this precise overlap is not obvious, since in general, isospectral potentials can admit much stronger deviations from each other.
\par
Finally, we want to comment on the role of rotation. Astrophysical black holes are expected to spin significantly, especially the observable ones forming via binary black hole mergers. This introduces not only a second parameter to the problem, but also complicates the structure of the metric and the perturbation equations. Dealing with this circumstance appropriately is far beyond the scope of this work, but an important aspect in the full problem. Note that the RZ metric has been extended to include rotation in \cite{Konoplya:2016jvv}, so our study can in principle be extended in the future. Since we find some degeneracy between the leading RZ parameters and the black hole mass, the situation might be significantly more difficult if the mass and spin are both treated as free parameters.
\section{Conclusions}\label{conclusions}
Using the spectrum of quasi-normal modes to study alternative black hole space-times is an interesting and feasible approach in exploring alternative theories of gravity. In a theory agnostic situation, where the underlying field equations are not known, parametrized metrics can be an effective way to investigate different properties of space-time. Unfortunately, the absence of the field equations makes it impossible to predict the evolution of gravitational perturbations, which are accessible to the LIGO and Virgo gravitational wave observatories.
\par
In this work, we assumed that some classes of alternative theories exist, in which non-rotating black hole solutions are sufficiently well approximated by the leading order terms of the parametrized metric proposed by Rezzolla and Zhidenko \cite{PhysRevD.90.084009}. Since scalar perturbations evolve qualitatively similar to the ones for gravitational perturbations in general relativity, and do not directly depend on the field equations, we have motivated that at least for some alternative theories, their qualitative behavior might be well captured with the scalar test field. Although no stringent direct connection to gravitational perturbations is possible, such fields are a legitimate approach if one is aware of this limitation.
\par
By using the third order WKB formula to compute the quasi-normal mode spectrum, we have investigated some part of the multi-dimensional parameter space of the RZ metric with reasonable resolution. We showed explicitly how the leading parameters influence the fundamental mode $(n=0)$ and the first overtone ($n=1$) for different $l$. We find that the parameter combinations $[a_0, b_0]$ and $[a_1,b_1]$ strongly correlate with the black hole mass $M$. This implies, even in the rather optimistic case in which the connection to the scalar field holds, tiny uncertainties in the black hole mass predict non-zero parameters of the metric, which are still in agreement with the pure Schwarzschild modes.
\par
\begin{figure*}[ht]
	\includegraphics[width=0.45\linewidth]{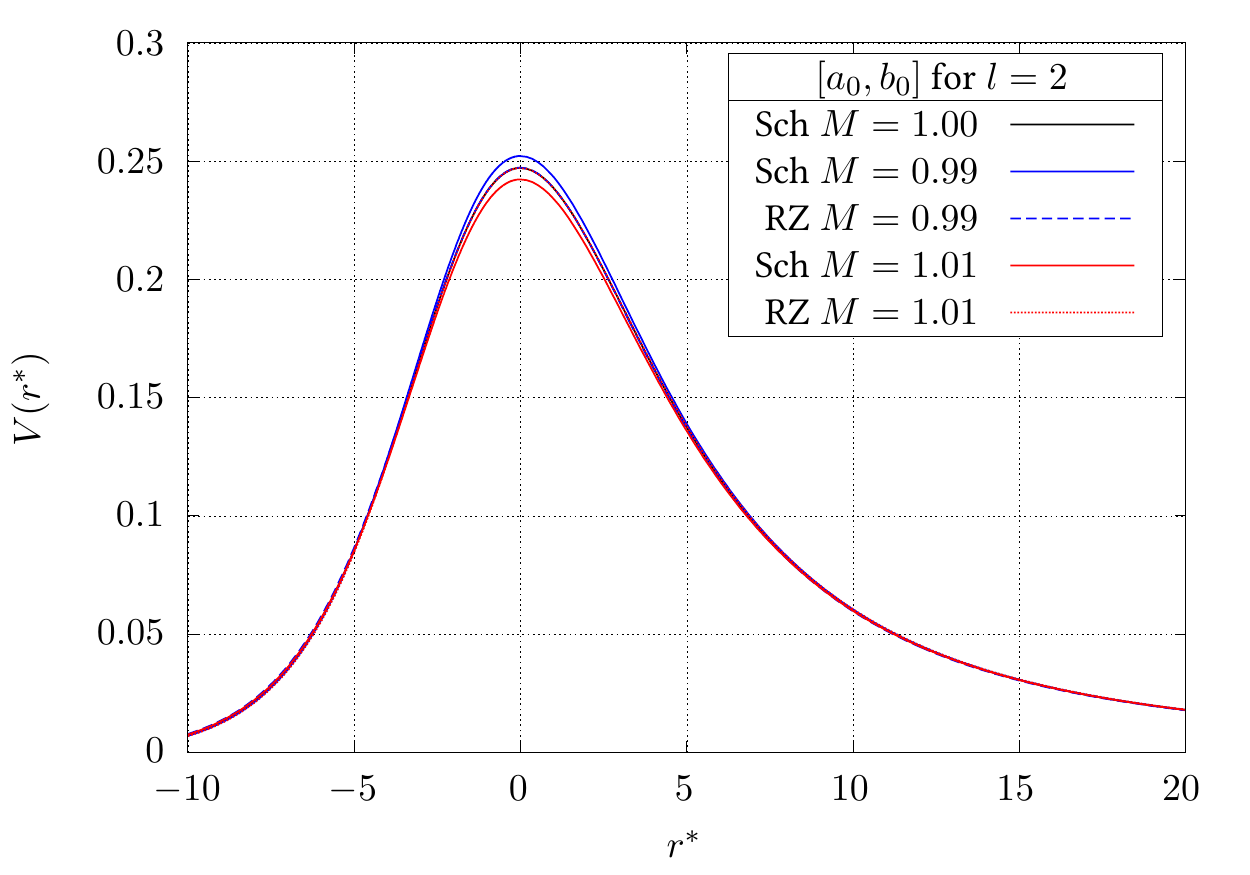}
	\includegraphics[width=0.45\linewidth]{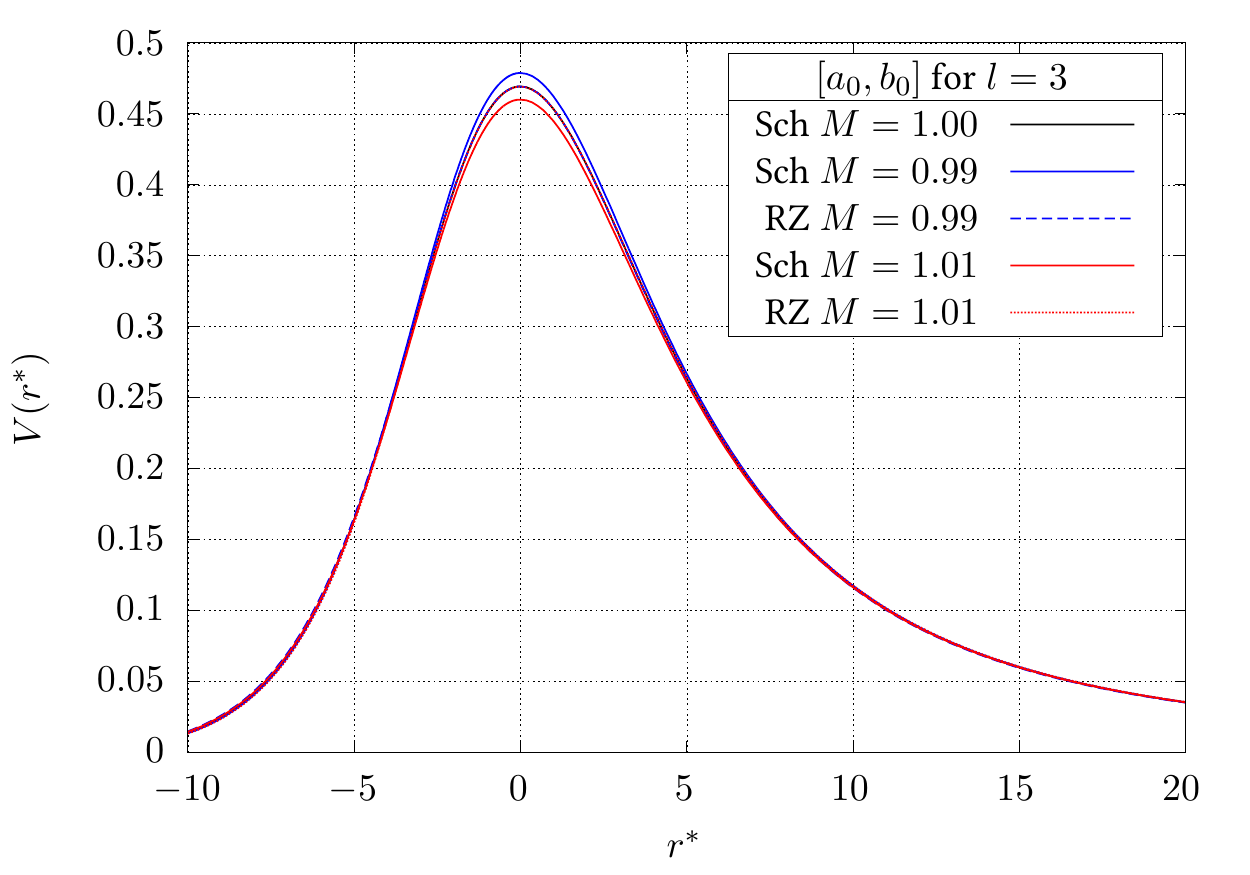}
	\\
	\includegraphics[width=0.45\linewidth]{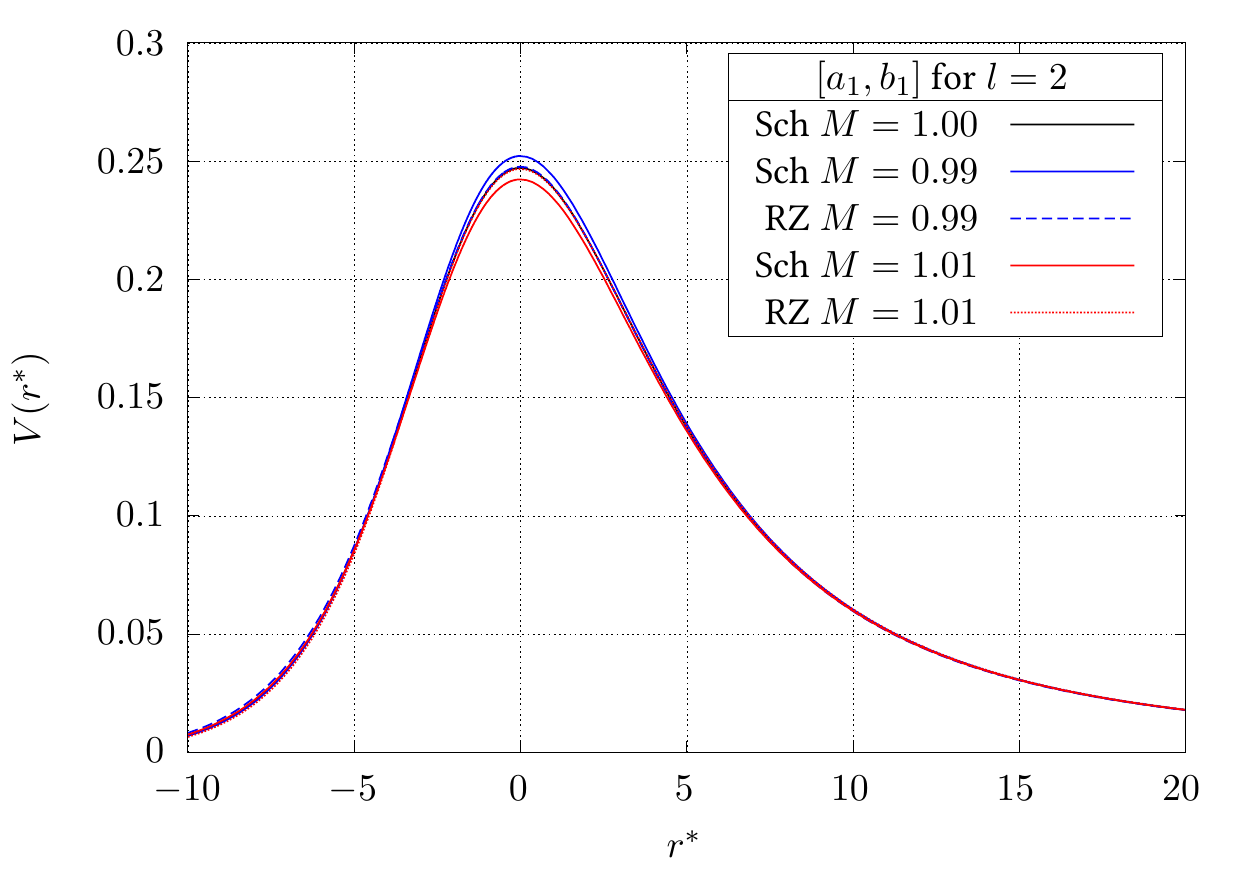}
	\includegraphics[width=0.45\linewidth]{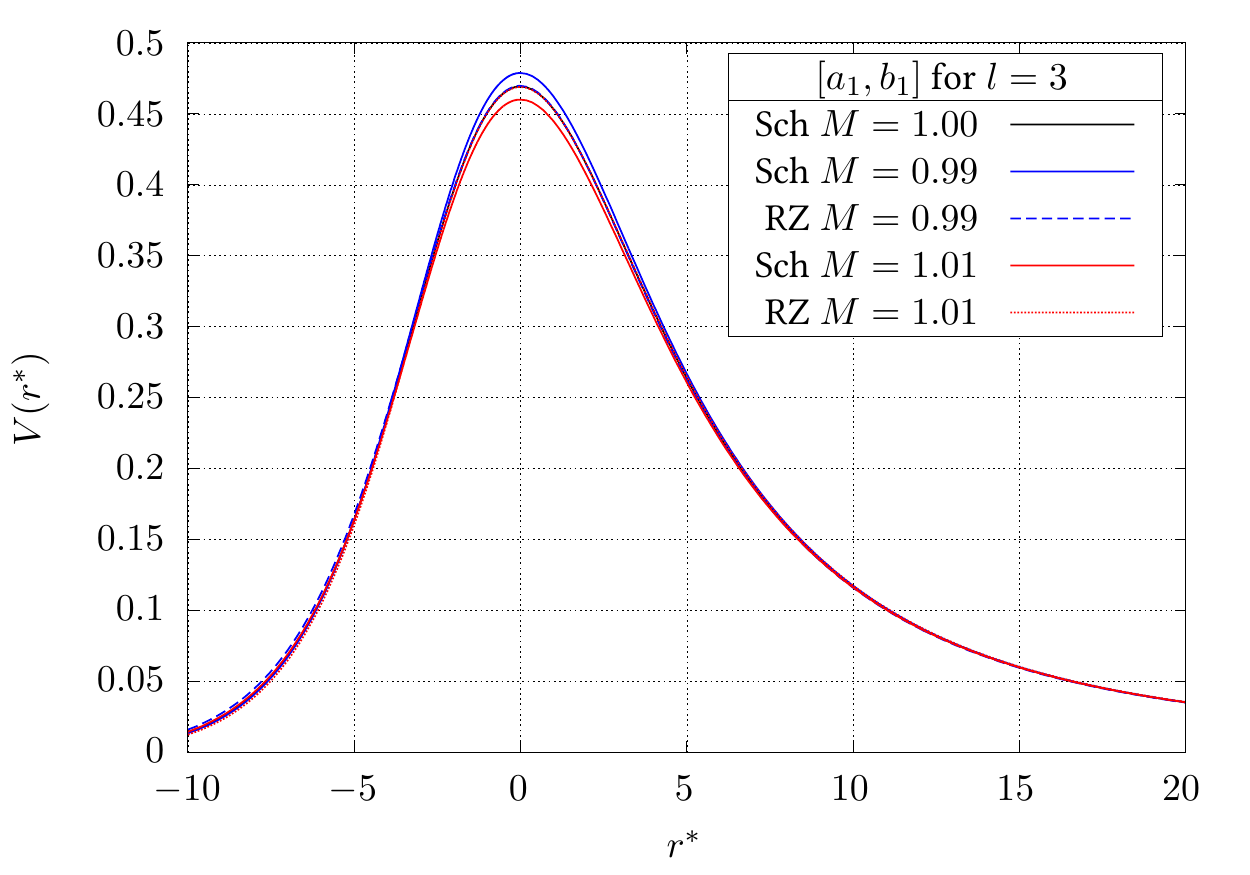}
	\caption{Here we show various potential barriers $V(r)$ as function of the tortoise coordinate $r^*$ for different parameters $[M, a_0, b_0]$ (top panels) and $[M, a_1, b_1]$ (bottom panels). From left to right we consider input from $n=0$ for the $l=2$ and $l=3$ cases. In each panel we show the standard Schwarzschild potentials (Sch) for different $M=[0.99, 1.00, 1.01]$ (solid lines). The dashed lines are the best matches of $[a_0, b_0]$ and $[a_1, b_1]$ for $M=[0.99, 1.01]$ in the RZ metric (dashed lines). The parameters were chosen by taking the smallest combined error $\delta \omega_{c}$ with respect to the $M=1$ Schwarzschild case. To demonstrate that these modified masses change the Schwarzschild potential noticeably, we also show the Schwarzschild potentials for $M=[0.99, 1.01]$ (red and blue solid) for comparison. Note that the RZ results (dashed lines) overlap with the canonic Schwarzschild potential (black solid) around the maximum, while the other Schwarzschild cases (red and blue solid) are clearly distinguishable.\label{potential}}
\end{figure*}

As we have demonstrated in Fig. \ref{potential}, the approach carried out in this work can also be used in the inverse spectrum problem, where one tries to reconstruct perturbation potentials from the quasi-normal mode spectrum. Our explicit reconstruction of potentials admitting the same fundamental modes proofs that such potentials can be constrained via this method.
\par
Our findings might put the popular approach of using the quasi-normal mode spectrum in the study of black hole space-times in a delicate context. Of course, in principle the higher order parameters might be well constrained once the whole spectrum is being provided with pristine precision, but such a scenario is far away from actual observations. In the loudest event GW150914, only rough constraints on the fundamental quasi-normal mode are possible \cite{LIGO1,LIGO2,Carullo:2019flw}. Since we did not consider rotating black holes in this work, our work has to be extended in the future. However, besides having a second parameter to fit observations, the results of this work already indicate that the situation becomes more involved and the unique reconstruction of parameterized space-times complicated.
\par
Finally, since parametrized metrics are also used to obtain geodesics and therefore find applications in the calculation of black hole shadows and ray tracing, it might be promising to combine both approaches. Since the lightring is closer to black hole horizon than typical extensions of an accretion disc, it might be enlightening to combine both approaches to put even tighter constraints on the parameters.
%\\~\\
\acknowledgments
S.V. thanks Andreas Boden and Sourabh Nampalliwar for useful discussions. S.V. receives the PhD scholarship Landesgraduiertenf\"orderung. \\The authors acknowledge support from the COST Action GW-verse CA16104.
\\~\\~\\
\bibliography{literatur1}
\appendix

\end{document}